\pdfoutput=1
\documentclass[11pt]{article}
\usepackage[]{acl}

\usepackage{color}
\usepackage{bbm}
\usepackage{multirow}
\usepackage[inline]{enumitem}
\usepackage{graphicx}
\usepackage{subfigure} 
\usepackage{sistyle}
\usepackage[ruled]{algorithm2e}
\usepackage{booktabs}
\usepackage{caption}
\SIthousandsep{,}
\usepackage{makecell}
\usepackage[skip=0pt]{caption}
\usepackage{amsmath}
\usepackage{hyperref}
\usepackage{array} 
\usepackage{acronym}
\usepackage[export]{adjustbox}

\newcommand{\trec}{TREC DL\xspace}
\newcommand{\nq}{NQ\xspace}
\newcommand{\webap}{WebAP\xspace}
\newcommand{\modelname}{ITEM\xspace}
\newcommand{\gti}{GTI-NQ\xspace}
\usepackage{times}
\usepackage{latexsym}
\usepackage[T1]{fontenc}
\usepackage[utf8]{inputenc}
\usepackage{microtype}
\usepackage{inconsolata}
\usepackage{graphicx} 
\usepackage{xcolor}
\usepackage{soul}
\newcommand{\heading}[1]{\vspace*{0.5mm}\noindent\textbf{#1.}}

\AtBeginDocument{%
  \providecommand\BibTeX{{%
    \normalfont B\kern-0.5em{\scshape i\kern-0.25em b}\kern-0.8em\TeX}}}

\makeatletter
\g@addto@macro\normalsize{%
  \abovedisplayskip 2pt plus1pt 
  \belowdisplayskip 2pt plus1pt
  \abovedisplayshortskip  2pt plus1pt%
  \belowdisplayshortskip  1pt plus1pt
}
\makeatother

\author{Hengran Zhang \textsuperscript{\rm 1,2,3}\space\space 
Keping Bi\textsuperscript{\rm 1,2,3}\footnotemark[1]\space\space 
Jiafeng Guo\textsuperscript{\rm 1,2,3}\footnotemark[1]\space\space 
\textbf{Xueqi Cheng}\textsuperscript{\rm 1,2,3}\space\space\\
\textsuperscript{\rm 1} State Key Laboratory of AI Safety \\
\textsuperscript{\rm 2} Institute of Computing Technology, Chinese Academy of Sciences \\
\textsuperscript{\rm 3}University of Chinese Academy of Sciences \\ 
\{zhanghengran22z, bikeping, guojiafeng, cxq\}@ict.ac.cn
 }

\title{An Iterative Utility Judgment Framework Inspired by Philosophical Relevance via LLMs}

\author{Hengran Zhang \textsuperscript{\rm 1,2,3}\space\space 
Keping Bi\textsuperscript{\rm 1,2,3}\footnotemark[1]\space\space 
Jiafeng Guo\textsuperscript{\rm 1,2,3}\footnotemark[1]\space\space 
\textbf{Xueqi Cheng}\textsuperscript{\rm 1,2,3}\space\space\\
\textsuperscript{\rm 1} State Key Laboratory of AI Safety \\
\textsuperscript{\rm 2} Institute of Computing Technology, Chinese Academy of Sciences \\
\textsuperscript{\rm 3}University of Chinese Academy of Sciences \\ 
\{zhanghengran22z, bikeping, guojiafeng, cxq\}@ict.ac.cn
 }

\begin{document}
\maketitle
\begin{abstract}
Relevance and utility are two frequently used measures to evaluate the effectiveness of an information retrieval (IR) system. 
Relevance emphasizes the aboutness of a result to a query,  while utility refers to the result's usefulness or value to an information seeker. 
In retrieval-augmented generation (RAG), high-utility results should be prioritized to feed to LLMs due to their limited input bandwidth. 
Re-examining RAG's three core components—relevance ranking derived from retrieval models, utility judgments, and answer generation—aligns with Schutz’s philosophical system of relevances, which encompasses three types of relevance representing different levels of human cognition that enhance each other.
These three RAG components also reflect three cognitive levels for LLMs in question-answering. 
Therefore, we propose an Iterative utiliTy judgmEnt fraMework (\modelname) to promote each step in RAG. 
We conducted extensive experiments on retrieval (\trec, \webap), utility judgment task (\gti), and factoid question-answering (\nq) datasets. 
Experimental results demonstrate improvements of \modelname in utility judgments, ranking, and answer generation upon representative baselines. Our code and benchmark can be found at \url{https://github.com/Trustworthy-Information-Access/ITEM}. 
\end{abstract}
\footnotetext[1]{Corresponding authors}
\renewcommand{\thefootnote}{\arabic{footnote}}
\setcounter{footnote}{0}
\section{Introduction}

\begin{figure}[t]
    \centering
    \includegraphics[width=\linewidth]{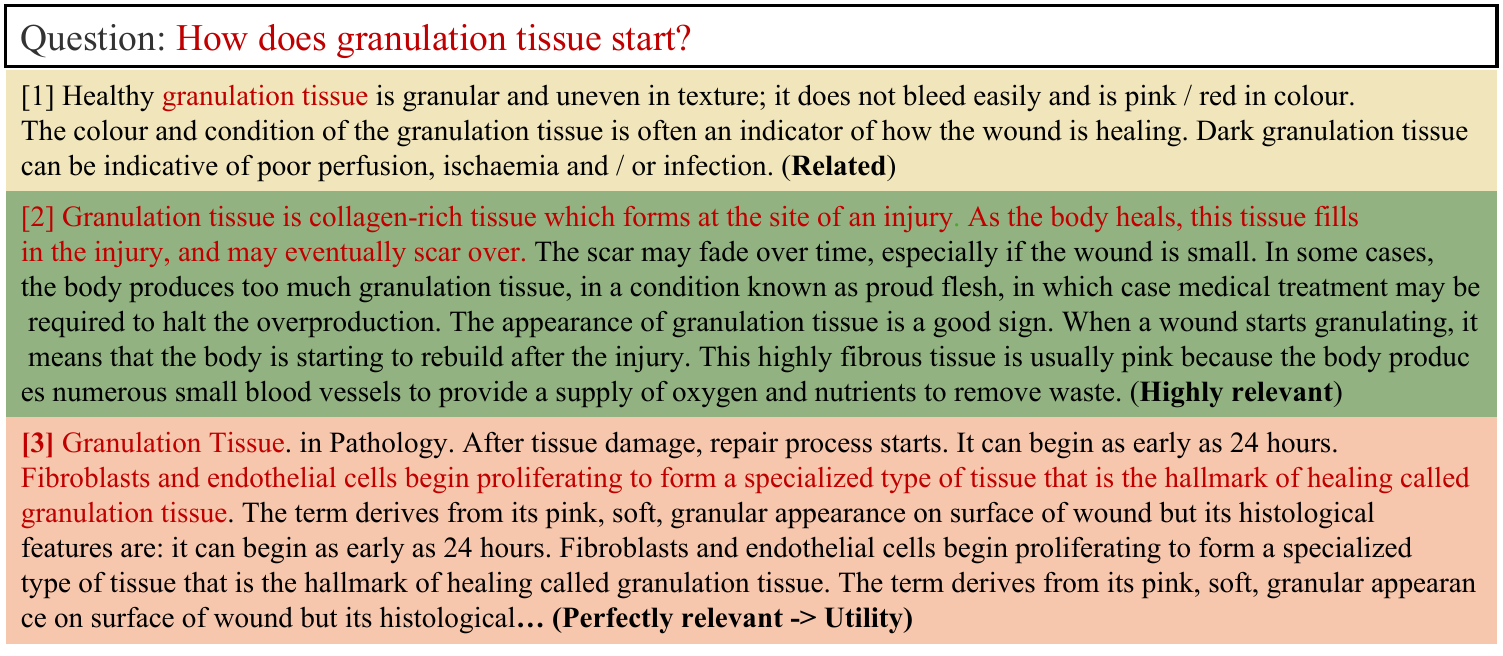}
    \caption{An example between utility and relevance from \trec dataset.}
    \label{fig:utility_relevance}
\end{figure}

Relevance and utility are two frequently used measures to evaluate Information Retrieval (IR) performance \cite{saracevic1996relevance, saracevic1975relevance, saracevic1988study}. 
Relevance usually refers to \textit{topical relevance} that measures the degree of fit between the subjects of a query and retrieved items, and the criteria of ``aboutness'' is used \cite{saracevic1988study, schamber1988relevance}. 
In contrast, 
\textit{utility} refers to the ``usefulness'' of retrieval items to an information seeker, of which the criterion is their ``value'' to the user \cite{saracevic1996relevance, saracevic1988study}. 
As the example from the \trec dataset shown in Figure \ref{fig:utility_relevance}, 
topical relevance does not necessarily mean utility, while utility indicates a higher standard of relevance. 
Since topical relevance is relatively easy to observe and measure \cite{schamber1990re}, the studies of IR models have been primarily focused on improving relevance for a long time \cite{bruce1994cognitive}. 

In the modern LLM era, retrieval-augmented generation (RAG) has become a hot research topic that equips LLMs with external knowledge \cite{xie2023adaptive, shi2023replug, izacard2023atlas, su2024dragin, glass2022re2g}.
Given the constrained bandwidth of LLM inputs, it is essential to prioritize high-value results to guide LLMs. 
Consequently, utility needs to be emphasized more than topical relevance in RAG. More recently, \citet{Zhang2024AreLL} highlighted the use of LLMs for utility judgments. 
In this paper, we aim to further promote the utility judgment performance of LLMs so that RAG can be enhanced by high-utility references.

\begin{figure}[t]
    \centering
    \includegraphics[width=\linewidth]{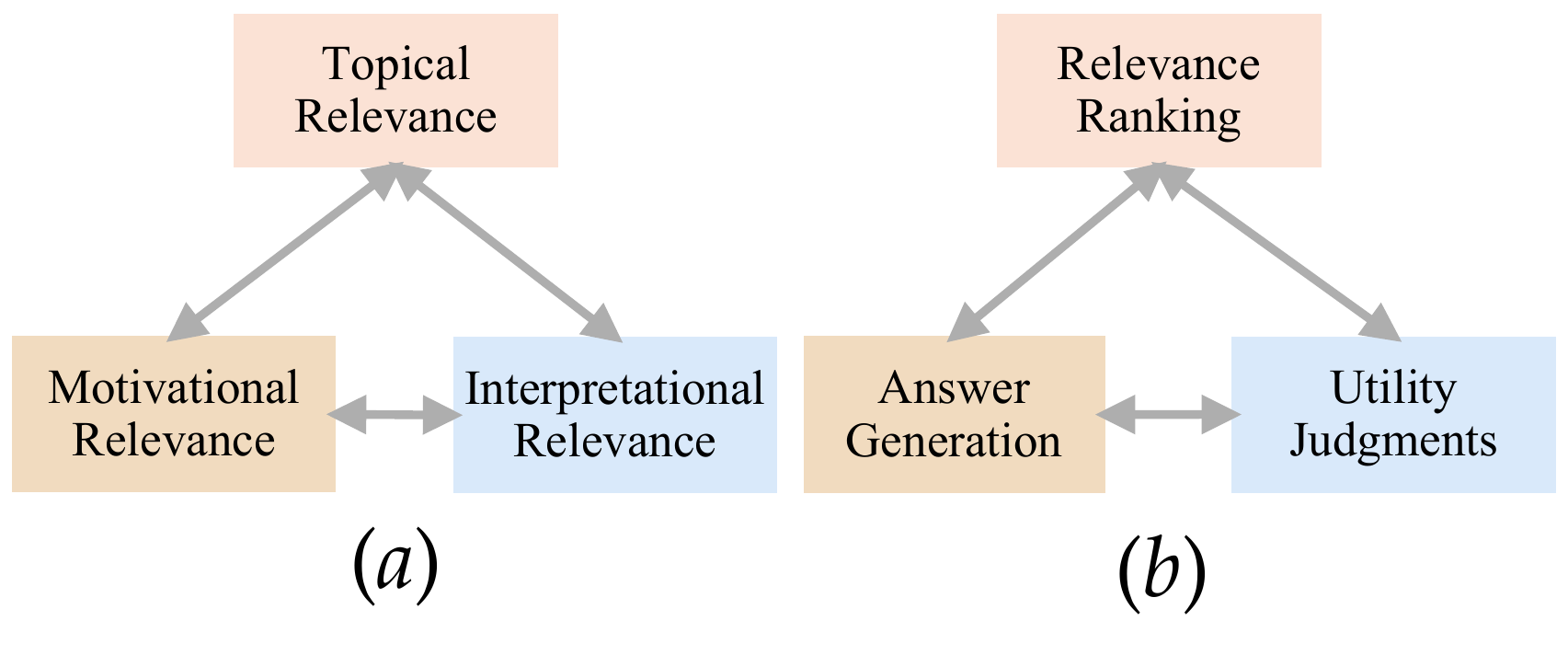}
    \caption{(a): Schutz's ``system of relevancies'', (b): the relation of each relevance to the components in RAG. The same color in the two frameworks is the corresponding connection.}
    \label{fig:shutz_rag}
\end{figure}

\heading{Schutz's Philosophical Theory of Relevance} 
Relevance is foundational in information retrieval (IR) and remains widely debated \cite{mizzaro1998many}. 
\citet{saracevic1996relevance} discussed the nature of relevance in the IR system as the effectiveness of interactive exchange on different levels, and they are non-independent interdependencies, which are primarily influenced by Schutz's philosophical theory of relevance. 
Schutz considered relevance as the property that determines the connections and relations in our lifeworld. 
He identified three types of basic and interdependent relevance that interact dynamically within a ``system of relevancies'' \cite{saracevic1996relevance, schutz2011reflections}:
\begin{enumerate*}[label=(\roman*)]
    \item Topical relevance, which refers to the perception of what is separated from one's experience to form one's present object of concentration;
    \item Interpretational relevance, which involves the past experiences in understanding the currently concerned object; and  
    \item Motivational relevance, which refers to the course of action to be adapted based on the interpretations.
\end{enumerate*}
The motivational relevance, in turn, helps obtain additional materials to become a user's new experience, which further facilitates topical and interpretational relevance.
Schutz posited that one's perception of the world may be enhanced by this dynamic interaction, as shown in Figure \ref{fig:shutz_rag}. 
By incorporating utility judgments into RAG, we can re-examine its three components: topical relevance or relevance ranking derived from retrieval models, utility judgments, and answer generation. 
Topical relevance is an emerging focus on a topic, utility is the deeper understanding of the topic, and answers indicate the final solution based on the interpretations and will guide users' actions. 
\textit{Therefore, topic relevance, utility, and derived answer also reflect three cognitive levels for LLMs in question-answering, from low to high, i.e., aboutness, the value for deriving an answer, and the derived answer.}

\heading{Iterative utiliTy judgmEnt fraMework (\modelname)}
Inspired by the philosophical theory of relevance, we believe the dynamic interactions between the three components in RAG can promote the performance of each step.
To verify the idea, we leverage LLMs to perform each step in RAG shown in Figure \ref{fig:shutz_rag}, i.e., relevance ranking, utility judgments (classification), and answer generation. 
We propose an Iterative utiliTy judgmEnt fraMework (\modelname) framework to enhance the utility judgment and QA performance of LLMs by interactions between the steps. 
\modelname has two variants depending on whether relevance ranking is involved in the iterations. We are curious to see which option will be better for the tasks: fewer iterations with more components in an iteration, more iterations with fewer components in an iteration, or more iterations with more components.

We experiment on various information-seeking tasks, i.e., multi-grade passage retrieval on TREC DL \cite{craswell2020overview}, multi-grade non-factoid answer passage retrieval on WebAP \cite{yang2016beyond}, utility judgments benchmark on \gti \cite{Zhang2024AreLL}, and factoid QA on NQ \cite{kwiatkowski2019natural}.
Experimental results have demonstrated that \modelname can outperform competitive baselines, including various single-shot judgment approaches in terms of utility judgments, topical relevance ranking, and answer generation, which confirms the viability of the adaptation of Schutz’s viewpoint of the relevance system into RAG. 
We also find that: 1) for difficult tasks (i.e., utility judgments of non-factoid answer passages in WebAP)  and complicated candidate passage list (i.e.,  \gti), more components in the iteration and multiple iterations are usually more beneficial; 2) our \modelname achieves performance comparable to the long reasoning mode while requiring very lower computational cost, thereby offering a more efficient and practical solution for evidence refinement; 3) for factoid QA tasks, more iterations with fewer components performs the best, indicating that more components and more iterations are not always needed, especially for simpler tasks.

\section{Related Work} 
\heading{Multi-dimensional relevance} 
The concept of ``relevance'' is central to information retrieval theory. Researchers have extensively debated its definition and measurement \cite{mizzaro1997relevance}. 
Early approaches primarily defined and assessed relevance through exact term matching \cite{vickery1959structure} or logical entailment \cite{hillman1964notion}. 
However, subsequent empirical studies revealed the limitations of system-oriented relevance analysis, prompting diverse perspectives on relevance \cite{saracevic1975relevance, swanson1986subjective, saracevic1996relevance, lancaster1968information, goffman1964methodology, kemp1974relevance, bruce1994cognitive}. 
For example, \citet{cooper1971definition} introduced logical relevance and utility. 
\citet{saracevic1996relevance} summarized five frameworks for information science: systems, communication, situational, psychological, and interaction frameworks, and categorized five distinct types of relevance, i.e., 1) system or algorithmic relevance; 2) topical or subject relevance; 3) cognitive relevance or pertinence; 4) situational relevance or utility; and 5) motivational or affective relevance. 
\citet{bruce1994cognitive} explored cognitive dimensions of relevance. 
Over time, scholarly consensus has coalesced around two primary perspectives: the system view and user view, with topical relevance and utility serving as their respective representative frameworks.

\heading{Utility-Focused Information Retrieval}
Utility is a distinct measure of relevance compared to topical relevance \cite{zhao2024beyond, saracevic1988study, saracevic1975relevance, saracevic1996relevance, Kaixin2024characterizing, zhang2023relevance}, and more recently, \citet{Zhang2024AreLL} highlighted the use of LLMs for utility judgments.
However, \citet{Zhang2024AreLL} only conducted a preliminary exploration of LLMs in utility judgments. 
Our work aims to further explore how to improve the performance of utility judgments for LLMs.

\heading{Retrieval-Augmented Generation (RAG)} 
RAG approaches are widely employed to mitigate the hallucination issues in large language models (LLMs) \cite{xie2023adaptive, zhou2024metacognitive, su2024dragin}. 
The current RAG approaches are categorized as follows:
\begin{enumerate*}[label=(\roman*)]
    \item single-round retrieval \cite{borgeaud2022improving, lewis2020retrieval, glass2022re2g, izacard2023atlas, shi2023replug}, which involves using the initial input as a query to retrieve information from an external corpus and then the information is incorporated as part of the input for the model; and 
    \item multi-round retrieval \cite{su2024dragin, jiang2023active, ram2023context, khandelwal2019generalization, trivedi2022interleaving}, which needs multi-round retrieval based on feedback from LLMs.  
\end{enumerate*} 

\heading{Iterative Relevance Feedback via LLMs}
Recent works \cite{li2023llatrieval, shao-etal-2023-enhancing} have achieved great success in using LLMs to obtain the information needs of the question as pseudo-relevance feedback for iterative retrieval.
They posit that a single retrieval may not yield comprehensive information, thus requiring multiple retrievals.
In contrast, our methodology involves making iterative utility judgments on the results obtained from a single retrieval. 

\section{Utility Judgments (UJ) via LLMs} 
Drawing inspiration from Schutz’s framework, we re-examine the three core components of RAG—namely, relevance ranking from retrieval models, utility judgments from the selector, and answer generation—and observe that they closely align with Schutz’s system. 
Specifically, topical relevance, utility, and the generated answer correspond to three cognitive levels in LLM-based question answering: aboutness, the value for answer derivation, and the final answer itself, respectively, moving from lower to higher cognitive processing. 
Inspired by Schutz's theory, we presume they can also interact with each other and enhance each other. 
Therefore, we propose an Iterative uTility-focused Evidence refineMent (\modelname) framework for utility judgments. 
\vspace{-1mm}
\subsection{Utility Judgments Definition}
Following \citet{Zhang2024AreLL}, given a question $q$ and a list of retrieved passages  $\mathcal{D} = [p_1, p_2, ..., p_n]$,  the goal of utility judgments for LLMs is to identify a set of passages $U=\{p_1, ...,p_m\}$, $m$ is the number of passage with utility selected by LLMs. 
There are two typical input approaches for LLMs to select the set $U$ from $\mathcal{D}$: pointwise and listwise. 
The pointwise approach independently evaluates the utility of individual passages as a binary classification task, while the listwise method assesses multiple passages simultaneously using the entire list as input. 

\vspace{-1mm}
\subsection{Single-Shot Utility Judgments}
\label{single-shot-utility}
The most common approach to judge utility for the $LLM$ is to perform a single-shot utility judgment, i.e., $U=LLM(q, \mathcal{D}, I)$, where $I$ is the instruction.
To improve the accuracy of utility judgments, \citet{Zhang2024AreLL} proposed a unified framework where the LLM generates a pseudo-answer $a$ and performs the utility judgments in a single output: $a,U=LLM(q, \mathcal{D}, I)$, i.e., \textbf{UJ-ExpA} in our baselines.  
\begin{figure}[t]
    \centering
    \includegraphics[width=\linewidth]{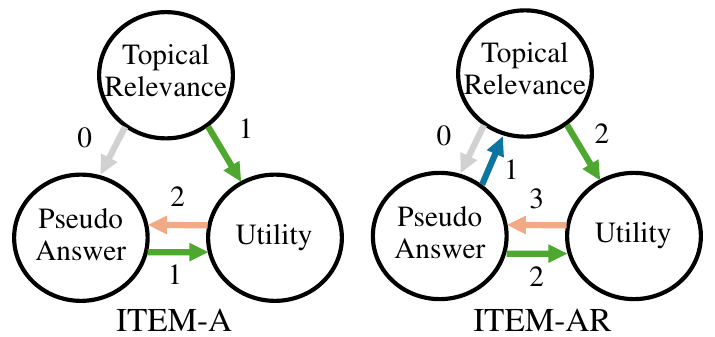}
    \caption{
    Flowchart illustrating the first iteration of \modelname. For \modelname-A, the process involves pseudo-answer generation followed by utility judgments and pseudo-answer generation. For \modelname-AR, the process includes pseudo-answer generation, relevance re-ranking, utility judgments, and pseudo-answer generation. 
 }
    \label{fig:framework}
\end{figure}

\vspace{-1mm}
\begin{figure}[t]
    \centering
    \includegraphics[width=\linewidth]{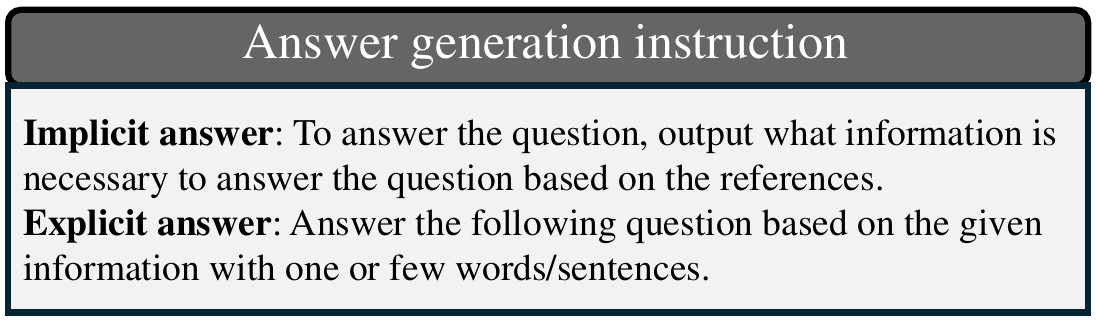}
    \vspace{-3mm}
    \caption{$I_a$ instruction contains the \textit{implicit answer} and \textit{explicit answer}.}
\vspace{-1mm}
    \label{fig:answer_instructions}
\end{figure}
\begin{figure}[t]
    \centering
    \includegraphics[width=\linewidth]{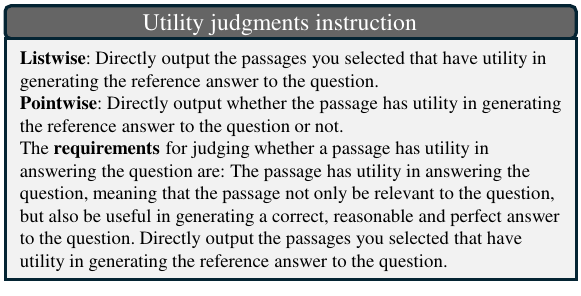}
    \vspace{-3mm}
    \caption{$I_u$ instruction contains listwise and pointwise approaches.}
\vspace{-1mm}
    \label{fig:utility_judgments_instructions}
\end{figure}

\subsection{Iterative utiliTy judgmEnt fraMework (\modelname)}
Drawing on Schutz’s theory of relevance, we introduce an Iterative utiliTy judgmEnt fraMework (\modelname) for RAG. The framework iteratively refines evidence through cycles of topical relevance ranking, pseudo-answer generation, and utility judgments. To manage LLM inference cost, two variants are proposed: \modelname‑A and \modelname‑AR, which implement iterative loops between two or three RAG components, respectively, as shown in Figure \ref{fig:framework}.

\heading{\modelname with Answering in the Loop (\modelname-A)}
Formally, at each iteration $t$ ($t \geq 1$), given the pseudo answer $a_t$  generated based on the utility judgment result $U_{t-1}$ from the previous iteration, we perform utility judgments on the candidate passages list $\mathcal{D}$ to obtain a set of passages with utility $U_t$: 
\begin{align}
    a_t &= LLM(q, U_{t-1}, I_a), \\
    U_t &= LLM(q, \mathcal{D}, a_t, I_u),
\end{align} 
where $I_a$ represents the answer prompts for LLMs (as detailed in Figure \ref{fig:answer_instructions}),  
$a_t$ can be in two forms: 
\begin{enumerate*}[label=(\roman*)]
    \item \textit{explicit answer} to the question $q$;
    \item \textit{implicit answer} that specifies the necessary information to answer the question $q$. 
\end{enumerate*} 
$I_u$ denotes the utility judgment prompts for LLMs (as detailed in Figure \ref{fig:utility_judgments_instructions}). 
We consider $U_0 = \mathcal{D}$ as the initial candidate set, where  $\mathcal{D}$ represents the initial results ranked by a retriever such as BM25 \cite{robertson2009probabilistic}.

\begin{figure}[t]
    \centering
    \includegraphics[width=\linewidth]{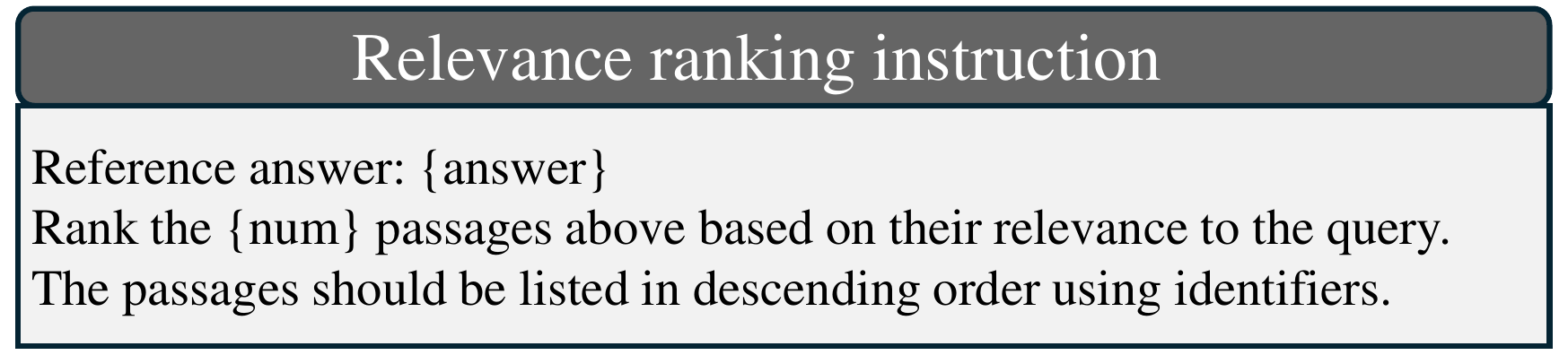}
    \vspace{-2mm}
    \caption{$I_r$ instruction}
    \label{fig:ranking_instructions}
\end{figure}

\heading{\modelname with both Answering and Ranking of Topical Relevance in the Loop (\modelname-AR)}
In the \modelname-A framework, topical relevance is not updated during the iteration process. 
To incorporate dynamic updating of topical relevance, we integrate a relevance ranking task into the \modelname framework, ensuring that all three tasks are executed in a loop. 
Formally, at iteration $t$ ($t \geq 1$), the answer $a_t$ is generated based on the judging result $U_{t-1}$ from the previous iteration. 
Subsequently, given $a_t$, the passage list $R_{t-1}$ from the previous iteration is ranked based on the relevance to the question, yielding a new ranked list $R_t$. 
Finally, the judging result $U_t$ is derived using the ranked list $R_t$ and the answer $a_t$: 
\begin{align}
    a_t &= LLM(q, U_{t-1}, I_a), \\
    R_t &= LLM(q, R_{t-1}, a_t, I_r), \\
    U_t &= LLM(q, R_t, a_t, I_u), 
\end{align} 
where  $I_r$ is the relevance ranking prompt for LLMs (as detailed in Figure \ref{fig:ranking_instructions}), respectively.

\heading{Overall}
At iteration $t$, we have two ways to produce the set $U_t$:
\begin{enumerate*}[label=(\roman*)]
    \item \texttt{Set-based approach}: Asking LLMs to identify the set of passages that have utility using listwise and pointwise input forms, which called \modelname-A$_s$ or \modelname-AR$_s$ variants; 
    \item \texttt{Rank-based approach}: Requesting LLMs to provide a ranked passage list based on utility (utility ranking prompt is shown in Appendix \ref{app:sec:ranking}) using the listwise input approach and considering the top-$k$ passages in the list to build $U_t$, which called \modelname-A$_r$ or \modelname-AR$_r$ variants.
    We set $k=5$ and more details of $k$ are shown in Appendix \ref{app:p_values}. 
    We find that \modelname-AR$_r$ does not improve ranking performance as well as \modelname-A$_r$ (see Appendix \ref{app:item-arr} for experimental analysis), so we do not employ \modelname-AR$_r$ in the ranking experiment. 
\end{enumerate*}
The rank-based approach has poor performance on the utility judgment task (details can be found in Appendix \ref{app:item-arr}), so we only employ the set-based approach on the utility judgment task. 
We stop the iteration when at most $m$ ($m$=3 in our paper) iterations are reached or the set of selected passages does not change, i.e., $t = m$ or $U_t = U_{t-1}$. 
Full details of all prompts can be found in Appendix  \ref{app:instruction}.

\vspace{-3mm}
\section{Experimental Setup}
\subsection{Datasets}
Our experiments are conducted on four benchmark datasets, including two retrieval datasets, i.e., \trec \cite{craswell2020overview} and \webap \cite{yang2016beyond}, a utility judgment dataset, i.e., \gti \cite{Zhang2024AreLL}, and an open-domain question answer (ODQA) dataset, i.e., NQ \cite{kwiatkowski2019natural}. Detailed statistics of the experimental datasets are shown in Table Appendix \ref{app:dataset}. We use two representative retrievers to gather candidate passages in $\mathcal{D}$ for utility judgments on \trec, \webap, and \nq datasets. 
Construction details can be found in Appendix \ref{app:retrievers}. 

\heading{\trec}  
We use the TREC-DL19 and TREC-DL20 datasets \cite{craswell2020overview}. 
Judgments of \trec are on a four-point scale, i.e., ``perfectly relevant'', ``highly relevant'', ``related'', and ``irrelevant''. 
We consider the passages that are ``perfectly relevant'' to have utility. 
We filter questions of two datasets that contain the passages labeled ``perfectly relevant: The passage is dedicated to the query and contains the exact answer. '' and combine them to form a whole dataset, i.e., the \trec. The annotation instruction for different points is shown in Appendix \ref{app:dataset} from the original paper. From the annotation instruction, we can find that the highest level of relevance (i.e., perfectly relevant: The passage is dedicated to the query and contains the exact answer.) in the TREC DL dataset can be seen as aligning with the definition of utility; therefore, we use passages labeled with the highest relevance level in TREC DL as positive examples for utility judgments evaluation. 

\heading{\webap} WebAP \cite{yang2016beyond} is a non-factoid answer passage collection built on Gov2. 
Non-factoid questions usually require longer answers, such as sentence-level or passage-level \cite{keikha2014evaluating, yang2016beyond, keikha2014retrieving}.
Relevant passages are annotated and categorized as ``perfect'', ``excel'', ``good'', and ``fair''. The annotation instruction is similar to TREC DL. 
So we considered the ``perfect'' passages to have utility.

\heading{\nq} Natural Questions (\nq) consist of factoid questions issued to the Google search engine \cite{kwiatkowski2019natural}. Each question is annotated with a long answer (typically a paragraph) and a short answer (one or more entities). 
Following \citet{Zhang2024AreLL}, we use the questions that have long answers in our experiments. 

\heading{\gti} Ground-truth inclusion (GTI) benchmark is constructed by \citet{Zhang2024AreLL} for utility judgment  task. 
The \gti constructs a candidate passage set of 10 passages for each query sourced from the NQ dataset, comprising the long answer (designated as the utility passage), highly relevant noisy passages, weakly relevant noisy passages, and counterfactual passages. 



\vspace{-2mm}
\subsection{Evaluation metrics}
For the utility judgments task, we evaluate the results of judgments using \textbf{P}recision, \textbf{R}ecall, and \textbf{F1}.
For the ranking task, we use the normalized discounted cumulative gain (\textbf{NDCG}) \cite{jarvelin2017ir} to evaluate the ranking performance. 
For the answer generation task, we use the standard exact match (EM) metric and F1.

\vspace{-2mm}
\subsection{LLMs}
We conduct our experiments using several representative LLMs, i.e., 
\begin{enumerate*}[label=(\roman*)]
    \item ChatGPT \cite{chatgpt} (we use the gpt-3.5-turbo-1106 version),
    \item Mistral \cite{jiang2023mistral} (the Mistral-7B-Instruct-v0.2 version), and 
    \item Llama 3 \cite{llama3} (the Meta-Llama-3-8B-Instruct version). 
    \item Qwen 3 \cite{yang2025qwen3} (Qwen3-8B version), supporting seamless switching between thinking mode and non-thinking mode within a single model. 
\end{enumerate*}
To ensure the reproducibility of the experiments, the temperature for all experiments is set to 0.

\begin{table*}[t]
\small
  \centering
   \setlength\tabcolsep{3.5pt}
    \begin{tabular}{l ccc ccc ccc ccc}
    \toprule
    \multirow{3}[6]{*}{Method} &   \multicolumn{6}{c}{WebAP}  &  \multicolumn{6}{c}{TREC DL}  \\
\cmidrule(r){2-7}  \cmidrule(r){8-13}          & \multicolumn{3}{c}{Listwise} & \multicolumn{3}{c}{Pointwise} & \multicolumn{3}{c}{Listwise} & \multicolumn{3}{c}{Pointwise}   \\
\cmidrule(r){2-4}  \cmidrule(r){5-7}   \cmidrule(r){8-10}  \cmidrule(r){11-13}    
& M & L& C & M & L& C & M & L& C & M & L& C \\       
    \midrule
    Vanilla & 20.79  & 21.79 & 28.43  & 23.05  & 25.09  & 26.85  & 45.67 & 49.39 & 55.19  & 45.11  & 47.64  & 49.84   \\
    
    UJ-ExpA  & 27.94  & 26.99 & 30.50  & 25.27  & 29.25  & 27.44 & 54.10 & 52.83 & 57.49  & 43.53  & 53.73  & 48.09  \\
    UJ-ImpA  & 25.06  & 26.22 & 29.89  & 28.35  & 25.29  & 26.32  & 48.29 & 48.22 & 56.18  & 48.31  & 50.20  & 48.83  \\
    $5$-sampling &  30.16 & 28.97 & 31.49   &   -    &    -   &   -   & 52.31 & 52.68 & 60.49  &    -   &  -     &    -     \\ 
    \midrule
    \modelname-A$_s$  $w.$ ExpA (1) & 29.76  & 27.50 & \underline{36.89}  & \underline{29.10}  & \underline{31.08}  & \underline{32.02} & \underline{53.78} & \underline{53.66} & \underline{62.52}  & \underline{49.44}  & \underline{52.09}  & 53.61    \\
    \modelname-A$_s$ $w.$ ImpA (1) & 26.06  & 25.59 & 34.97  & 28.28  & 30.53  & 29.34 & 49.39 & 53.73 & 58.11  & 46.01  & 53.68  & \underline{54.61}    \\
     \modelname-AR$_s$ $w.$ ExpA(1) & \underline{35.50}  & \underline{\textbf{31.44}} & 36.58  &  -     &   -    &  -     & 52.34 & 48.97 & 62.00  &     -  &  -     &    -    \\
     \midrule
    \modelname-A$_s$  $w.$ ExpA (3) & 31.65  & \underline{29.32} & 39.57  & \underline{\textbf{30.50}} & \underline{\textbf{32.67}} & 31.43  & 54.86 & \underline{\textbf{56.03}} & \underline{\textbf{63.18}} & \underline{\textbf{51.74}} & 52.46  & \underline{\textbf{55.74}}  \\
    \modelname-A$_s$ $w.$ ImpA (3) & 28.36  & 26.10 & \underline{\textbf{40.78}} & 30.13  & 29.64  & \underline{\textbf{32.54}} & 52.05 & 55.14 & 60.56  & 46.59  & \underline{\textbf{53.76}} & 54.90  \\ 

    \modelname-AR$_s$ $w.$ ExpA(3) & \underline{\textbf{37.06}}  & 29.08 & 38.58  &   -    &  -     &   -    & \underline{\textbf{56.27}} & 52.10 & 61.37  &   -    &  -     &   -    \\

    \bottomrule
    \end{tabular}%
    \caption{The F1 performance (\%) of utility judgments with different LLMs on the different datasets (the numbers in parentheses represent $m$-values). ``-'' indicates no experiments are performed under the pointwise approach because of that the $k$-sampling method and our \modelname-AR$_s$ require listwise input. \textbf{bold} indicates the best performance. \underline{Underline} means the best performance among all variants of our \modelname with the same $m$ value. ``M'', ``L'', and ``C'' mean ``Mistral'', ``Llama 3'' and ``ChatGPT'', respectively.}
  \label{tab:retrieval_exp}%
\end{table*}%

\subsection{Baselines}
We utilize the following baselines on the utility judgments task and question answering performance based on the utility judgment results: 

   \heading{Single-shot utility judgments}
    \begin{enumerate*}[label=(\roman*)]
         \item \textbf{Vanilla}: Ask LLMs to provide utility judgments based on the instruction directly.
        
        \item \textbf{UJ-ExpA}: Utility judgments and provide explicit answers simultaneously through a single output, which is shown to be effective in \citet{Zhang2024AreLL}. 
        
        \item \textbf{UJ-ImpA}:  Utility judgments and provide implicit answers that are necessary to answer the question through a single output.  
    \end{enumerate*}

    \heading{$k$-sampling} \cite{Zhang2024AreLL} proposed $k$-sampling to alleviate the sensitivity of LLMs to input order. Specifically, the $k$-sampling method randomizes the order of the input passage list $k$ times in addition to the original input and aggregates the $k+1$ utility judgment results through voting. 
    For fair comparison, we use the $k=5$, more details are in Appendix \ref{app:k-sampling}.

    To evaluate the effectiveness of the proposed \modelname framework in ranking tasks,  we are using a verbalized ranking. 
    Therefore, we also employ another verbalized ranking method, i.e., \textbf{RankGPT} \cite{sun2023chatgpt} as our main baseline, which uses the LLMs to directly rank input passages based on their relevance to the query. 
\begin{table}[t]
\small
  \centering
   \setlength\tabcolsep{1.5pt}
    \begin{tabular}{lcccc}
    \toprule
    \multirow{2}[4]{*}{Method} & \multicolumn{2}{c}{Llama3-8B} & \multicolumn{2}{c}{ChatGPT} \\
\cmidrule{2-5}          & Listwise & Pointwise & Listwise & Pointwise \\
    \midrule
    Vanilla & 43.38  & 28.55  & 59.37  & 35.31  \\
    UJ-ExapA & 47.07  & 39.32  & 66.13  & 37.17  \\
    UJ-ImpA & 43.31  & 38.72  & 57.40  & 37.29  \\
    k-sampling & 49.20  & -     & 71.17  & - \\
    \midrule
    ITEM-As-ExpA (1) & 49.26  & \underline{47.52}  & 72.44  & \underline{54.89} \\
    ITEM-As-Imp (1) & 47.47  & 37.98  & 68.92  & 43.17 \\
    ITEM-ARs-ExpA (1) & \underline{50.77} & -     &  \underline{74.43}     & - \\
    \midrule
    ITEM-As-ExpA (3) & 49.73  & \textbf{48.90 } & 73.55  & \textbf{55.45} \\
    ITEM-As-Imp (3) & 48.03  & 38.34  & 69.68  & 43.58 \\
    ITEM-ARs-ExpA (3) & \textbf{51.22 } & -     &  \textbf{76.34}     & - \\
    \bottomrule
    \end{tabular}%
    \caption{The F1 performance (\%) of utility judgments with different LLMs on the GTI-NQ dataset. \textbf{Bold} and \underline{Underline} are defined in Table \ref{tab:retrieval_exp}.}
  \label{tab:gti-nq}%
\end{table}%
\begin{table}[t]
\small
  \centering
   \setlength\tabcolsep{2.5pt}
    \begin{tabular}{lcccc}
    \toprule
    \multirow{2}[4]{*}{Method} & \multicolumn{2}{c}{Qwen3} & \multicolumn{2}{c}{Time (s) / Query} \\
\cmidrule(r){2-3} \cmidrule(r){4-5} &  List  & Point  & List  & Point  \\
    \midrule
    Vanilla ($w/o$ Think)   & 41.82 &  42.30  & \phantom{1}0.4 & \phantom{1}0.7 \\
    UJ-ExapA ($w/o$ Think)  & 47.30 & 43.65  & \phantom{1}1.5 & \phantom{1}3.2 \\
    UJ-ImpA ($w/o$ Think)   & 43.96 & 38.92  & \phantom{1}0.9 & \phantom{1}1.6 \\
    k-sampling ($w/o$ Think)   & 49.28 & - & \phantom{1}9.2 & -\\
    \midrule
    Vanilla ($w.$ Think) & 55.96 & - & 31.2 & - \\
    \midrule
    ($m$=1, $w/o$ Think) \\
    ITEM-As-ExpA  & 50.53 & \underline{52.56}  & \phantom{1}2.0 & \phantom{1}4.3 \\ 
    ITEM-As-Imp & 51.58 & 52.43  & \phantom{1}1.4 & \phantom{1}1.9 \\
    ITEM-ARs-ExpA &  \underline{55.72}  & - & \phantom{1}3.3 & - \\
    \midrule
    ($m$=3, $w/o$ Think)\\
    ITEM-As-ExpA  & 51.69 & \textbf{53.33}  & \phantom{1}3.4 & 10.3 \\
    ITEM-As-Imp  & 52.38 & 52.34  & \phantom{1}2.4 & \phantom{1}3.6 \\
    ITEM-ARs-ExpA &  \textbf{56.02} & - & \phantom{1}7.1 & -\\
    \bottomrule
    \end{tabular}%
    \caption{The F1 performance (\%) of utility judgments with Qwen 3-8B on the GTI-NQ dataset. \textbf{Bold} and \underline{Underline} are defined in Table \ref{tab:retrieval_exp}. ``w/ thinking'' and ``w/o thinking'' refer to the model generating with its thinking function enabled and disabled, respectively. Due to inference cost constraints, our evaluation of the thinking function was conducted under the vanilla listwise input setting. The terms ``List'' and ``Point'' refer to the ``Listwise'' and ``Pointwise'' approaches.}
  \label{tab:gti-nq-qwen}%
\end{table}%
\vspace{-3mm}
\section{Experimental Results of LLMs}
This section will present the performance of each task within our \modelname framework. 
By default, the pseudo answer is the \textit{explicit answer} in all experiments, if not specified otherwise.

\subsection{Utility Judgment Results}
\label{retrieval} 
Table \ref{tab:retrieval_exp} shows the F1 performance on the \trec and \webap datasets using three LLMs.   
Further, we utilize a better-performing open-source LLM, i.e., Llama-3.1-8B, and a closed LLM, i.e., ChatGPT, to conduct experiments on \gti, as shown in Table \ref{tab:gti-nq}. 
We assess the long reasoning process of the popular LLM Qwen3 (Thinking) against baseline models and our own model without thinking, using the larger GTI-NQ dataset. Results are shown in Table \ref{tab:gti-nq-qwen}. 
Since \modelname-A$_r$ and  \modelname-AR$_r$ have poor F1 performance in utility judgments (refer to Table \ref{tab:iteration_rudif} in Appendix \ref{app:item-arr} for details), we restrict our experiments to \modelname-A$_s$ and \modelname-AR$_s$ in this section. 

\heading{\modelname with a Single Iteration vs. Baselines}
All LLMs using our \modelname with a single iteration generally outperform the single-shot utility judgments on three datasets and may even surpass the $k$-sampling method. 
For example, ChatGPT on the \trec dataset using our \modelname-A$_s$ $w.$ ExpA and ImpA in the listwise approach improve the F1 performance by 8.7\% and 3.4\% over UJ-ExpA and UJ-ImpA, respectively. 
Explicit generation of pseudo-answers by LLMs enhances their performance in utility judgment tasks, highlighting the importance of task interaction. 
Moreover, concurrent execution of answer generation and utility judgment within a single inference cycle yields inferior performance compared to sequential task execution through separate reasoning phases.  

\heading{\modelname with Multiple Iterations vs. \modelname with Single Iteration} 
All LLMs using our \modelname-A and \modelname-AR generally demonstrate improved performance with multiple iterations compared to single iterations on all three datasets. For instance, on the WebAP dataset, Mistral, Llama 3, and ChatGPT (using our \modelname-A $w.$ ExpA) improved their F1 scores in the listwise approach by 6.4\%, 6.6\%, and 7.3\%, respectively, after multiple iterations. 
Moreover, our method achieves state-of-the-art performance compared to all baselines by leveraging the iterative framework. 
The performance improvement from multiple iterations underscores the importance of iterative interaction and further supports Schutz’s interactive framework. 
Moreover, ChatGPT outperforms other LLMs on all datasets using both input approaches.

\heading{\modelname-A$_s$ vs. \modelname-AR$_s$} 
\label{item-a-vs-item-ar}
In our utility-emphasized iterative RAG framework, \modelname-A$_s$ and \modelname-AR$_s$ are the two major methods we propose. 
From Table \ref{tab:retrieval_exp}, we find that \modelname-AR$_s$ works better than \modelname-A$_s$ most of the time for complex questions (\webap, all the questions are non-factoid) and the complex candidate passage list (\gti, containing different types of passage), indicating complicated questions or passage lists need more components in the loop. 
For \trec, which contains factoid questions, we find that \modelname-AR$_s$ is worse than \modelname-A$_s$ most times on \trec. 
This is reasonable since factoid questions are relatively easier to answer and may not need more components involved in the iteration.

\heading{Listwise vs. Pointwise}  
The general performance of utility judgments for LLMs is better with the listwise approach than with the pointwise approach. 
The primary rationale lying in the listwise approach exposes the LLM to broader contextual information, thereby facilitating more effective interaction during the LLMs in judging the passages' utility. 

\heading{Thinking vs Non-thinking}
As shown in Table \ref{tab:gti-nq-qwen}, the long reasoning mode demonstrates superior performance compared to other single-shot baselines and the $k$-sampling method. 
Our proposed \modelname framework achieves comparable performance to this long reasoning method but at a significantly lower computational cost (only about 23\% of long reasoning). This indicates that the \modelname framework successfully balances efficiency with effectiveness, offering a more practical and resource-efficient solution for RAG. 
\begin{table}[t]
  \centering
  \small
   \setlength\tabcolsep{1pt}
    \begin{tabular}{ccccccc}
    \toprule
    \multirow{2}[4]{*}{Method} & \multicolumn{2}{c}{Mistral} & \multicolumn{2}{c}{Llama 3} & \multicolumn{2}{c}{ChatGPT} \\
\cmidrule(r){2-3} \cmidrule(r){4-5}  \cmidrule(r){6-7}          & TREC  & WebAP & TREC   & WebAP & TREC  & WebAP  \\
\midrule

    $D$  & 58.69  & 21.89  & 58.69  & 21.89  & 58.69  & 21.89  \\
    RankGPT & 69.81  & 29.34  & 75.61  & 41.73  & 80.56  & 42.49  \\
    \midrule
    \modelname-A$_r$ (1)  & 70.57  & 37.11  & 73.95  & 40.89    & 80.79   &  \underline{50.30}  \\
    \modelname-AR$_s$ (1) & \underline{71.29}  & \underline{37.48}  & \underline{77.22}  & \underline{43.80}  & \underline{81.38}  & 48.42  \\
    \midrule
    \modelname-A$_r$ (3) & \underline{\textbf{74.27}}  &  43.80 &   \underline{\textbf{77.34}} &  \underline{\textbf{45.88}}  &  \underline{\textbf{83.12}}  &  \underline{\textbf{51.61}}  \\
    \modelname-AR$_s$ (3) & 73.24  & \underline{\textbf{45.45}} & 74.80  & 44.87  & 82.89  &  48.80 \\

    \midrule
    \end{tabular}%
    \caption{
    The NDCG@5 performance (\%) of the ranking using different LLMs on the different datasets.  \textbf{Bold} and \underline{Underline}  are defined in Table \ref{tab:retrieval_exp}.
    }
 \label{tab:ranking_msitral}%
\end{table}%

\begin{table}[t]
  \centering
  \small
    \begin{tabular}{ccccc}
    \toprule
    \multirow{2}[4]{*}{Method} & \multicolumn{2}{c}{Llama 3} & \multicolumn{2}{c}{ChatGPT} \\
\cmidrule{2-5}          & @5    & @10   & @5    & @10 \\
    \midrule
    $D$ & 29.46  & 45.26  & 29.46  & 45.26  \\
    RankGPT & 71.50  & 74.05  & 77.27  & 78.64  \\
    \midrule
     ITEM-A$_r$ (1) & 74.36  & 76.91  & \underline{85.99}  & \underline{87.26}  \\
    ITEM-AR$_s$ (1) & \underline{75.46}  & \underline{77.75}  & 84.54  & 85.14  \\
    \midrule
    ITEM-A$_r$ (3) & 75.95  & 78.18  & \textbf{87.48}  & \textbf{88.47}  \\
    ITEM-AR$_s$ (3) & \textbf{76.38}  & \textbf{78.56}  & 85.95  & 86.39  \\
    \bottomrule
    \end{tabular}%
    \caption{ The NDCG performance (\%) of the ranking using different LLMs on the \gti dataset. \textbf{Bold} and \underline{Underline}  are defined in Table \ref{tab:retrieval_exp}.}
  \label{tab:gti-ranking}%
\end{table}%

\subsection{Ranking Performance} 
We also assess whether the ranking performance has been improved within \modelname on retrieval datasets (Table \ref{tab:ranking_msitral}) and utility judgments benchmark (Table \ref{tab:gti-ranking}). 
In terms of ranking performance, we consider two rankings: relevance ranking (\modelname-AR$_s$) and utility ranking (\modelname-A$_r$). 
We can observe that:
\begin{enumerate*}[label=(\roman*)]
    \item Our \modelname with a single iteration significantly improves the ranking of topical relevance performance compared to the RankGPT. 
    For instance, relevance ranking outperforms RankGPT in NDCG@5 by 2.1\% on the TREC dataset. 
    The performance improvement may stem from the interaction between tasks.
     \item After iterations, relevance and utility ranking performance have been improved on all datasets and all LLMs. 
     The ranking benefits from our dynamic iterative framework, confirming Schutz's theory of dynamic iterative interaction. 
     \item
     From Tables \ref{tab:ranking_msitral}\&\ref{tab:gti-ranking}, we can find that \modelname-AR$_s$  is generally better than \modelname-A$_r$ when $m=1$. 
     However, when $m=3$, it may have the opposite performance. 
     The possible reason is that when $m$ is small, the answer is not very good, and utility is more dependent on the answer than on relevance.  
     As iterations proceed, we observe improved answer quality and utility performance. In contrast, relevance does not show as marked an improvement. 
\end{enumerate*}


\subsection{Results of Answer Generation}
In the answer generation task, the results of utility judgments are fed to LLMs for answer generation. 
We use the factoid QA dataset (i.e., NQ)  for answer generation evaluation, as shown in Table \ref{tab:nq_exp}. 
From Tables \ref{tab:retrieval_exp}\&\ref{tab:gti-nq}, we find that the listwise approach generally outperforms the pointwise approach for utility judgments. 
Consequently, our answer generation experiments utilize only the listwise utility judgments. 
\begin{table}[t]
  \centering
  \small
   \vspace{-2mm}
    \begin{tabular}{ccccc}
    \toprule
    \multirow{2}[2]{*}{References} & \multicolumn{2}{c}{Mistral} & \multicolumn{2}{c}{ChatGPT} \\
\cmidrule(r){2-3} \cmidrule(r){4-5}      & EM    & F1      & EM    & F1 \\
\midrule
    Golden & 46.09  & 62.59    & 66.40  & 76.86  \\
    \midrule
    $D$ & 31.58  & 47.69  & 46.54  & 57.00  \\
    Vanilla & 31.16  & 47.43   & 48.52  & 58.64  \\
    UJ-ExpA & 32.76  & 48.46  & 47.72  & 58.01  \\
    UJ-ImpA & 30.67  & 46.83   & 49.01  & 59.30  \\
    5-sampling & 33.24  & 48.84     & 48.90  & 58.97  \\
    \midrule
    \modelname-A$_s$ (1) & 32.98  & 49.00     & 49.38  & \underline{59.78}  \\
    \modelname-AR$_s$ (1) & \underline{33.30}  & \underline{49.26}    & \underline{49.52}  & 59.64  \\
   \midrule
    \modelname-A$_s$ (3) & \underline{\textbf{33.73}} & \underline{\textbf{49.63}}   & \underline{\textbf{49.69}} & \underline{\textbf{60.18}} \\
    \modelname-AR$_s$ (3) & 33.40  & 49.27   & 49.06  & 59.67  \\
    \bottomrule
    \end{tabular}%
    \caption{The answer generation performance (\%) of all LLMs on the NQ dataset using reference passages collected from different methods. \textbf{Bold} means the best performance except for the answer generation with golden evidence. \underline{Underline} is defined in Table \ref{tab:retrieval_exp}.}
  \label{tab:nq_exp}%
\end{table}%
The following observations can be made from Table \ref{tab:nq_exp}:
\begin{enumerate*}[label=(\roman*)]
    \item \modelname outperforms baselines across all metrics on all LLMs (except for the EM score of Llama 3), indicating that \modelname can help the LLMs to find better evidence for generating answers. 
    \item  Similar to Tables \ref{tab:retrieval_exp}\&\ref{tab:gti-nq}, when the $m=1$, \modelname-AR$_s$ performs better than \modelname-A$_s$, which shows the importance of relevance reranking in \modelname. However, as the number of iterations increases, \modelname-A$_s$ performs better than \modelname-AR$_s$. 
     We are keen to discern the optimal choice for different tasks: 
     1) More components and more iterations are not always needed, especially for simpler tasks; 2) Fewer iterations with numerous components, or increased iterations with few components. 
\end{enumerate*}

\vspace{-2mm}
\section{Further Analyses}
\begin{figure}[t]
    \centering
    \includegraphics[width=\linewidth]{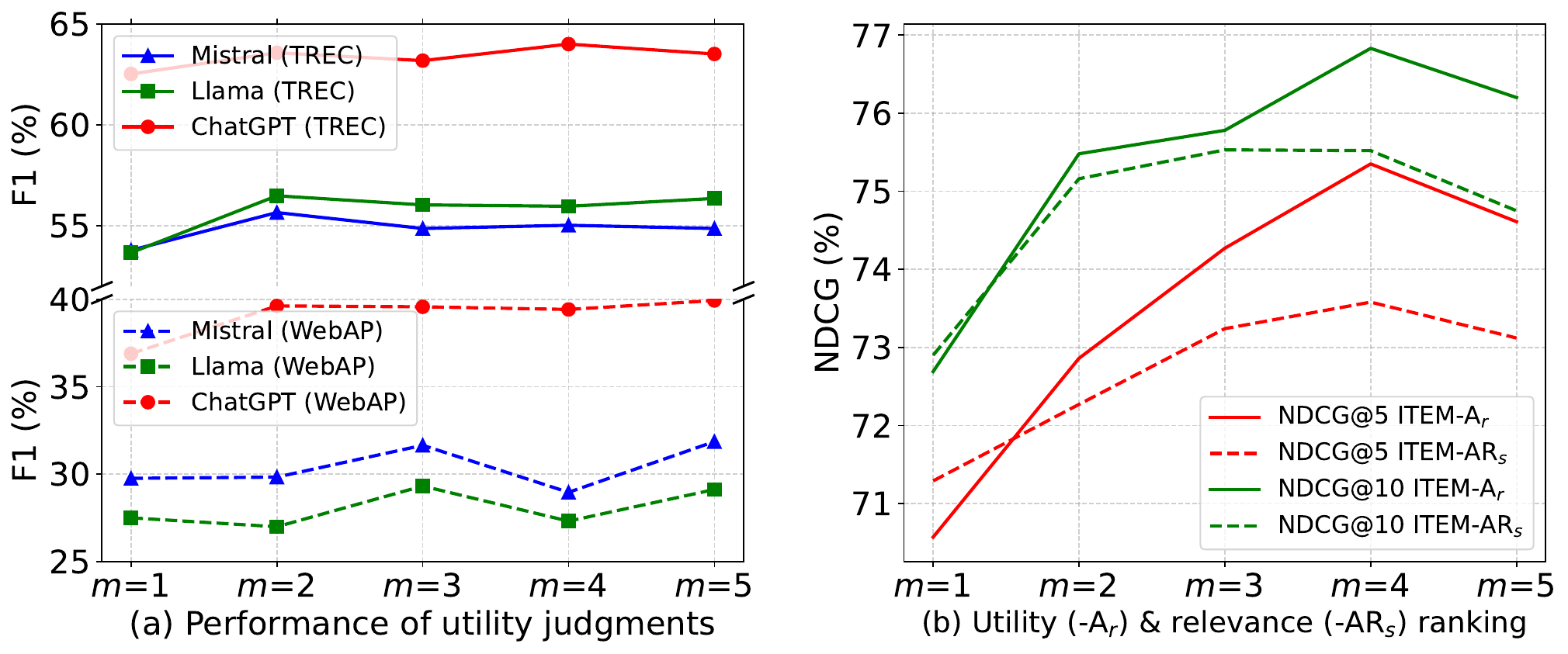}
    \caption{(a): utility judgments performance (\%) in terms of $m$ values in \modelname-A$_s$. (b): relevance ranking (\modelname-AR$_s$) and utility ranking (\modelname-A$_r$) performance (\%)  of Mistral on the \trec dataset. }
    \label{fig:ablation_utility_m}
\end{figure}

\begin{figure}[t]
    \centering
    \includegraphics[width=\linewidth]{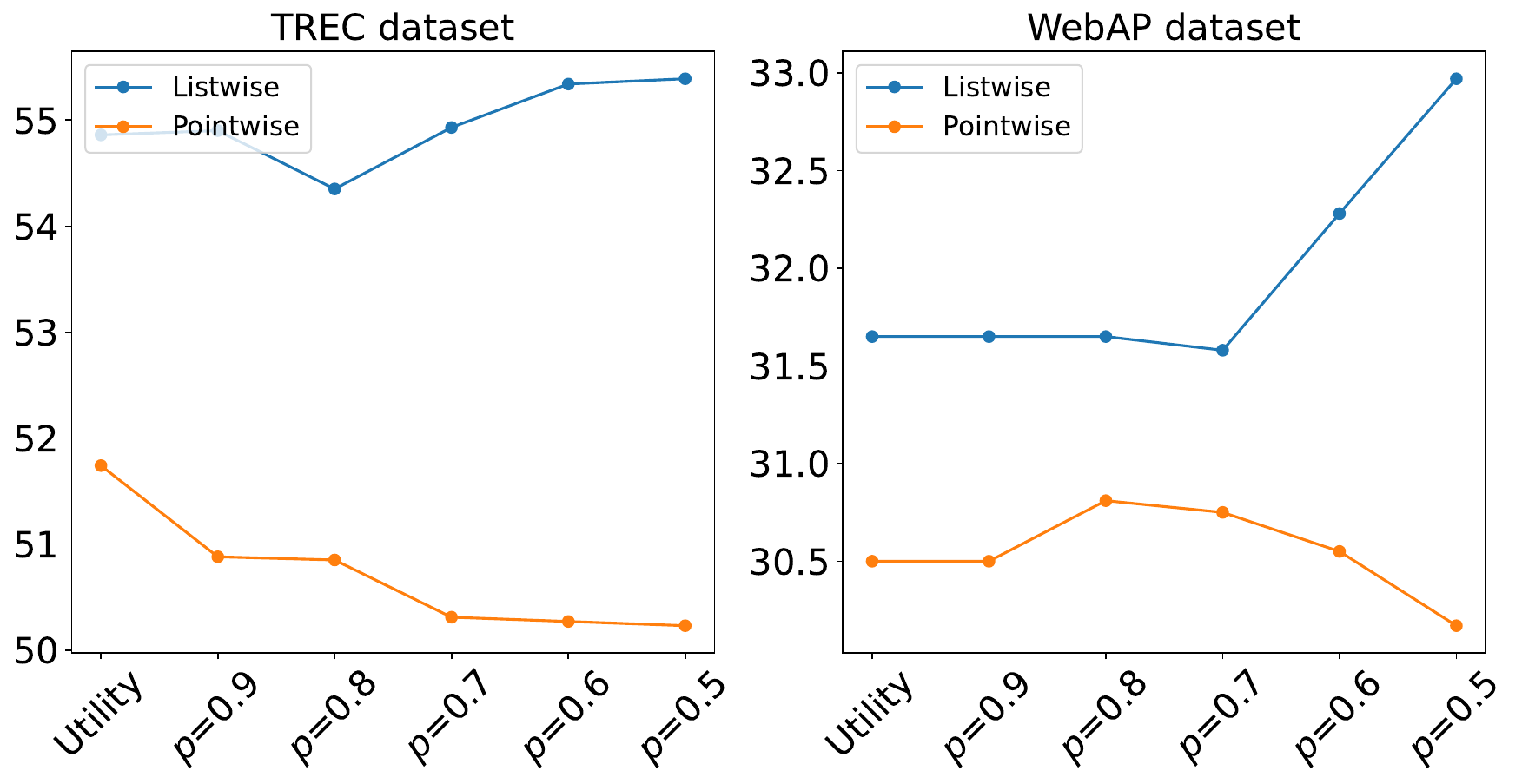}
    \caption{The utility judgments F1 performance (\%) of Mistral in different iteration stop conditions ($m$=3) under \modelname-A$_s$.}
    \label{fig:stop_conditions}
\end{figure}

\heading{Iteration Rounds}
\label{aba:m}
Figure \ref{fig:ablation_utility_m}  shows the performance of (a): utility judgments under \modelname-A$_s$  and (b): ranking with varying maximum iteration rounds $m$. 
We observe the following: 
1) Varying the value of $m$ affects the performance of utility judgments and ranking. 
2) 
Based on empirical observations balancing the cost and performance, $m$ was operationally configured with distinct values for different question types on utility judgments ($m$=3 in our paper on all experiments for fair comparison): $m$=2 for factoid questions, whereas $m$=3 is better for non-factoid questions in practical applications. 
3) Utility ranking generally outperforms relevance ranking, which confirms the effectiveness of utility in the ranking task. 
\heading{Iteration Stop Conditions}
In our experiment, we employ utility-based selection results as the stopping criterion—that is, the iteration halts when the selection outcomes remain identical across two consecutive rounds. 
In this section, we also evaluate an alternative stopping condition based on the pseudo-answer generation performance of \modelname. 
Specifically, we calculate the ROUGE-L score \cite{lin2004rouge} of the answer in two iterations and stop the iteration if the ROUGE-L of $a_t$ and $a_{t-1}$ is greater than $p$.  
The utility judgment performance of different iteration stop conditions is shown in Figure \ref{fig:stop_conditions}. 
The results show that using different stopping conditions affects the performance of utility judgments.
However, using the answer as a stopping condition, different LLMs on different datasets may need to look for different $p$, which is not very flexible.  

\vspace{-3mm}
\section{Conclusion}
In this paper, we propose an Iterative utiliTy judgmEnt fraMework (\modelname) to enhance the utility judgment and QA performance of LLMs by interactions between the steps, inspired by Schutz's philosophical discussion of relevance.   
This is a unified framework of iterative RAG with an emphasis on utility.  
Our framework achieves state-of-the-art performance in zero-shot scenarios, outperforming previous methods in utility judgments, ranking of topical relevance, and answer generation tasks, indicating that the cognitive process of LLMs on a specific topic can also be improved by a similar process. 
Our experiments also highlight the significance of dynamic interaction in achieving high performance and stability. 
Future directions include developing better fine-tuning strategies for utility judgments and creating end-to-end solutions for RAG.

\section*{Limitations}
There are two primary limitations that should be acknowledged: 
\begin{enumerate*}[label=(\roman*)]

    \item Our methods are applied in zero-shot scenarios without any training. 
    The zero-shot approach itself does not enhance the LLMs's inherent capability in utility judgments but rather employs strategies to improve performance on utility judgment tasks. 
    Future research should explore designing more effective training methods, e.g., utilizing our iterative framework with self-evolution techniques \cite{singh2023beyond}, to genuinely enhance the LLMs's ability in utility judgments through training.
    
    \item The number of candidate passages in the search scenario is much larger than 20. 
    The number of search results we assumed is too small. 
    We need to continue to study utility judgments in large-scale scenarios in the future. 
    
\end{enumerate*}

\section{Ethics Statement}
Our research does not rely on personally identifiable information. 
All datasets and models used in our paper are publicly available and have been widely adopted by researchers. 
We firmly believe in the principles of open research and the scientific value of reproducibility. 
To this end, we have made all data, and code associated with our paper publicly available on GitHub. 
This transparency not only facilitates the verification of our findings by the community but also encourages the application of our methods in other contexts. 


\section*{Acknowledgements}
This work was funded by the National Natural Science Foundation of China (NSFC) under Grants No. 62302486, No. 62441229, and the Innovation Project of ICT CAS under Grants No. E361140. 
%
\bibliography{custom}

@article{saracevic1975relevance,
  title={Relevance: A review of and a framework for the thinking on the notion in information science},
  author={Saracevic, Tefko},
  journal={Journal of the American Society for information science},
  volume={26},
  number={6},
  pages={321--343},
  year={1975},
  url={https://asistdl.onlinelibrary.wiley.com/doi/abs/10.1002/asi.4630260604},
  publisher={Wiley Online Library}
}

@inproceedings{jarvelin2017ir,
  title={IR evaluation methods for retrieving highly relevant documents},
  author={J{\"a}rvelin, Kalervo and Kek{\"a}l{\"a}inen, Jaana},
  booktitle={ACM SIGIR Forum},
  volume={51},
  number={2},
  pages={243--250},
  year={2017},
  url={https://dl.acm.org/doi/pdf/10.1145/345508.345545},
  organization={ACM New York, NY, USA}
}

@article{Zhang2024AreLL,
  title={Are Large Language Models Good at Utility Judgments?},
  author={Zhang, Hengran and Zhang, Ruqing and Guo, Jiafeng and de Rijke, Maarten and Fan, Yixing and Cheng, Xueqi},
  journal={Proceedings of the 47th International ACM SIGIR Conference on Research and Development in Information Retrieval (SIGIR '24)},
  year={2024},
  url={https://arxiv.org/pdf/2403.19216}
}

@article{zhao2024beyond,
  title={Beyond Relevance: Evaluate and Improve Retrievers on Perspective Awareness},
  author={Zhao, Xinran and Chen, Tong and Chen, Sihao and Zhang, Hongming and Wu, Tongshuang},
  journal={arXiv preprint arXiv:2405.02714},
  year={2024},
  url={https://arxiv.org/pdf/2405.02714}
}

@article{saracevic1988study,
  title={A study of information seeking and retrieving. I. Background and methodology},
  author={Saracevic, Tefko and Kantor, Paul and Chamis, Alice Y and Trivison, Donna},
  journal={Journal of the American Society for Information science},
  volume={39},
  number={3},
  pages={161--176},
  year={1988},
  publisher={Wiley Online Library},
  url={https://www.researchgate.net/publication/245088184_A_Study_in_Information_Seeking_and_Retrieving_I_Background_and_Methodology}
}

@inproceedings{saracevic1996relevance,
  title={Relevance reconsidered},
  author={Saracevic, Tefko},
  booktitle={Proceedings of the second conference on conceptions of library and information science (CoLIS 2)},
  pages={201--218},
  year={1996},
  url={}
}

@article{sun2023chatgpt,
  title={Is ChatGPT Good at Search? Investigating Large Language Models as Re-Ranking Agent},
  author={Sun, Weiwei and Yan, Lingyong and Ma, Xinyu and Ren, Pengjie and Yin, Dawei and Ren, Zhaochun},
  journal={EMNLP},
  year={2023},
  url={https://aclanthology.org/2023.emnlp-main.923.pdf}
}

@article{Kaixin2024characterizing,
    author = {Kaixin Ji and Danula Hettiachchi and Flora D. Salim and Falk Scholer and Damiano Spina},
    title = {Characterizing Information Seeking Processes with Multiple Physiological Signals},
    journal = {SIGIR},
    year = {2024},
    url={https://arxiv.org/pdf/2405.00322}
}

@inproceedings{xie2023adaptive,
  title={Adaptive chameleon or stubborn sloth: Revealing the behavior of large language models in knowledge conflicts},
  author={Xie, Jian and Zhang, Kai and Chen, Jiangjie and Lou, Renze and Su, Yu},
  booktitle={The Twelfth International Conference on Learning Representations},
  year={2023},
  url={https://arxiv.org/pdf/2305.13300}
}

@article{izacard2023atlas,
  title={Atlas: Few-shot learning with retrieval augmented language models},
  author={Izacard, Gautier and Lewis, Patrick and Lomeli, Maria and Hosseini, Lucas and Petroni, Fabio and Schick, Timo and Dwivedi-Yu, Jane and Joulin, Armand and Riedel, Sebastian and Grave, Edouard},
  journal={Journal of Machine Learning Research},
  volume={24},
  number={251},
  pages={1--43},
  year={2023},
  url={https://arxiv.org/pdf/2208.03299}
}

@inproceedings{zhou2024metacognitive,
    author = {Zhou, Yujia and Liu, Zheng and Jin, Jiajie and Nie, Jian-Yun and Dou, Zhicheng},
    title = {Metacognitive Retrieval-Augmented Large Language Models},
    year = {2024},
    isbn = {9798400701719},
    publisher = {Association for Computing Machinery},
    address = {New York, NY, USA},
    url = {https://doi.org/10.1145/3589334.3645481},
    doi = {10.1145/3589334.3645481},
    booktitle = {Proceedings of the ACM on Web Conference 2024},
    pages = {1453–1463},
    numpages = {11},
    location = {<conf-loc>, <city>Singapore</city>, <country>Singapore</country>, </conf-loc>},
    series = {WWW '24}
}

@article{su2024dragin,
  title={DRAGIN: Dynamic Retrieval Augmented Generation based on the Real-time Information Needs of Large Language Models},
  author={Su, Weihang and Tang, Yichen and Ai, Qingyao and Wu, Zhijing and Liu, Yiqun},
  journal={arXiv preprint arXiv:2403.10081},
  year={2024},
  url={https://arxiv.org/pdf/2403.10081}
}

@inproceedings{keikha2014evaluating,
  title={Evaluating answer passages using summarization measures},
  author={Keikha, Mostafa and Park, Jae Hyun and Croft, W Bruce},
  booktitle={Proceedings of the 37th international ACM SIGIR conference on Research \& development in information retrieval},
  pages={963--966},
  year={2014},
  url={https://dl.acm.org/doi/abs/10.1145/2600428.2609485}
}

@inproceedings{keikha2014retrieving,
  title={Retrieving passages and finding answers},
  author={Keikha, Mostafa and Park, Jae Hyun and Croft, W Bruce and Sanderson, Mark},
  booktitle={Proceedings of the 19th Australasian Document Computing Symposium},
  pages={81--84},
  year={2014},
  url={https://dl.acm.org/doi/10.1145/2682862.2682877}
}

@inproceedings{yang2016beyond,
  title={Beyond factoid QA: effective methods for non-factoid answer sentence retrieval},
  author={Yang, Liu and Ai, Qingyao and Spina, Damiano and Chen, Ruey-Cheng and Pang, Liang and Croft, W Bruce and Guo, Jiafeng and Scholer, Falk},
  booktitle={Advances in Information Retrieval: 38th European Conference on IR Research, ECIR 2016, Padua, Italy, March 20--23, 2016. Proceedings 38},
  pages={115--128},
  year={2016},
  organization={Springer},
  url={https://maroo.cs.umass.edu/pub/web/getpdf.php?id=1195}
}

@inproceedings{borgeaud2022improving,
  title={Improving language models by retrieving from trillions of tokens},
  author={Borgeaud, Sebastian and Mensch, Arthur and Hoffmann, Jordan and Cai, Trevor and Rutherford, Eliza and Millican, Katie and Van Den Driessche, George Bm and Lespiau, Jean-Baptiste and Damoc, Bogdan and Clark, Aidan and others},
  booktitle={International conference on machine learning},
  pages={2206--2240},
  year={2022},
  organization={PMLR},
  url={https://arxiv.org/pdf/2112.04426}
}

@article{lewis2020retrieval,
  title={Retrieval-augmented generation for knowledge-intensive nlp tasks},
  author={Lewis, Patrick and Perez, Ethan and Piktus, Aleksandra and Petroni, Fabio and Karpukhin, Vladimir and Goyal, Naman and K{\"u}ttler, Heinrich and Lewis, Mike and Yih, Wen-tau and Rockt{\"a}schel, Tim and others},
  journal={Advances in Neural Information Processing Systems},
  volume={33},
  pages={9459--9474},
  year={2020},
  url={https://dl.acm.org/doi/pdf/10.5555/3495724.3496517}
}

@inproceedings{glass2022re2g,
    title = "{R}e2{G}: Retrieve, Rerank, Generate",
    author = "Glass, Michael  and
      Rossiello, Gaetano  and
      Chowdhury, Md Faisal Mahbub  and
      Naik, Ankita  and
      Cai, Pengshan  and
      Gliozzo, Alfio",
    editor = "Carpuat, Marine  and
      de Marneffe, Marie-Catherine  and
      Meza Ruiz, Ivan Vladimir",
    booktitle = "Proceedings of the 2022 Conference of the North American Chapter of the Association for Computational Linguistics: Human Language Technologies",
    month = jul,
    year = "2022",
    address = "Seattle, United States",
    publisher = "Association for Computational Linguistics",
    url = "https://aclanthology.org/2022.naacl-main.194",
    doi = "10.18653/v1/2022.naacl-main.194",
    pages = "2701--2715",
  
}

@article{shi2023replug,
  title={Replug: Retrieval-augmented black-box language models},
  author={Shi, Weijia and Min, Sewon and Yasunaga, Michihiro and Seo, Minjoon and James, Rich and Lewis, Mike and Zettlemoyer, Luke and Yih, Wen-tau},
  journal={arXiv preprint arXiv:2301.12652},
  year={2023},
  url={https://arxiv.org/pdf/2301.12652}
}

@inproceedings{jiang2023active,
    title = "Active Retrieval Augmented Generation",
    author = "Jiang, Zhengbao  and
      Xu, Frank  and
      Gao, Luyu  and
      Sun, Zhiqing  and
      Liu, Qian  and
      Dwivedi-Yu, Jane  and
      Yang, Yiming  and
      Callan, Jamie  and
      Neubig, Graham",
    editor = "Bouamor, Houda  and
      Pino, Juan  and
      Bali, Kalika",
    booktitle = "Proceedings of the 2023 Conference on Empirical Methods in Natural Language Processing",
    month = dec,
    year = "2023",
    address = "Singapore",
    publisher = "Association for Computational Linguistics",
    url = "https://aclanthology.org/2023.emnlp-main.495",
    doi = "10.18653/v1/2023.emnlp-main.495",
    pages = "7969--7992",
}

@article{ram2023context,
  title={In-context retrieval-augmented language models},
  author={Ram, Ori and Levine, Yoav and Dalmedigos, Itay and Muhlgay, Dor and Shashua, Amnon and Leyton-Brown, Kevin and Shoham, Yoav},
  journal={Transactions of the Association for Computational Linguistics},
  volume={11},
  pages={1316--1331},
  year={2023},
  publisher={MIT Press One Broadway, 12th Floor, Cambridge, Massachusetts 02142, USA~…},
  url={https://aclanthology.org/2023.tacl-1.75.pdf}
}

@article{khandelwal2019generalization,
  title={Generalization through memorization: Nearest neighbor language models},
  author={Khandelwal, Urvashi and Levy, Omer and Jurafsky, Dan and Zettlemoyer, Luke and Lewis, Mike},
  journal={ICLR},
  url={https://arxiv.org/pdf/1911.00172},
  year={2020}
}

@inproceedings{trivedi2022interleaving,
  author       = {Harsh Trivedi and
                  Niranjan Balasubramanian and
                  Tushar Khot and
                  Ashish Sabharwal},
  editor       = {Anna Rogers and
                  Jordan L. Boyd{-}Graber and
                  Naoaki Okazaki},
  title        = {Interleaving Retrieval with Chain-of-Thought Reasoning for Knowledge-Intensive
                  Multi-Step Questions},
  booktitle    = {Proceedings of the 61st Annual Meeting of the Association for Computational
                  Linguistics (Volume 1: Long Papers), {ACL} 2023, Toronto, Canada,
                  July 9-14, 2023},
  pages        = {10014--10037},
  publisher    = {Association for Computational Linguistics},
  year         = {2023},
  url          = {https://doi.org/10.18653/v1/2023.acl-long.557},
  doi          = {10.18653/V1/2023.ACL-LONG.557},
  bibsource    = {dblp computer science bibliography, https://dblp.org}
}

@inproceedings{cohen2018wikipassageqa,
  title={WikiPassageQA: A benchmark collection for research on non-factoid answer passage retrieval},
  author={Cohen, Daniel and Yang, Liu and Croft, W Bruce},
  booktitle={The 41st international ACM SIGIR conference on research \& development in information retrieval},
  pages={1165--1168},
  year={2018},
  url={https://dl.acm.org/doi/pdf/10.1145/3209978.3210118}
}

@inproceedings{hashemi2020antique,
  title={ANTIQUE: A non-factoid question answering benchmark},
  author={Hashemi, Helia and Aliannejadi, Mohammad and Zamani, Hamed and Croft, W Bruce},
  booktitle={Advances in Information Retrieval: 42nd European Conference on IR Research, ECIR 2020, Lisbon, Portugal, April 14--17, 2020, Proceedings, Part II 42},
  pages={166--173},
  year={2020},
  organization={Springer},
  url={https://arxiv.org/pdf/1905.08957}
}

@inproceedings{hashemi2019performance,
  title={Performance prediction for non-factoid question answering},
  author={Hashemi, Helia and Zamani, Hamed and Croft, W Bruce},
  booktitle={Proceedings of the 2019 ACM SIGIR International Conference on Theory of Information Retrieval},
  pages={55--58},
  year={2019},
  url={https://dl.acm.org/doi/10.1145/3341981.3344249}
}

@inproceedings{bi2019iterative,
  title={Iterative relevance feedback for answer passage retrieval with passage-level semantic match},
  author={Bi, Keping and Ai, Qingyao and Croft, W Bruce},
  booktitle={Advances in Information Retrieval: 41st European Conference on IR Research, ECIR 2019, Cologne, Germany, April 14--18, 2019, Proceedings, Part I 41},
  pages={558--572},
  year={2019},
  organization={Springer},
  url={https://arxiv.org/pdf/1812.08870}
}

@inproceedings{voorhees2003overview,
  title={Overview of the TREC 2003 robust retrieval track.},
  author={Voorhees, Ellen M and others},
  booktitle={Trec},
  pages={69--77},
  year={2003}, 
  url={https://www.researchgate.net/publication/2912042_Overview_of_TREC_2003}
}

@article{craswell2020overview,
  title={Overview of the TREC 2019 deep learning track},
  author={Craswell, Nick and Mitra, Bhaskar and Yilmaz, Emine and Campos, Daniel and Voorhees, Ellen M},
  journal={arXiv preprint arXiv:2003.07820},
  year={2020},
  url={https://arxiv.org/pdf/2003.07820}
}

@article{kwiatkowski2019natural,
  title={Natural questions: a benchmark for question answering research},
  author={Kwiatkowski, Tom and Palomaki, Jennimaria and Redfield, Olivia and Collins, Michael and Parikh, Ankur and Alberti, Chris and Epstein, Danielle and Polosukhin, Illia and Devlin, Jacob and Lee, Kenton and others},
  journal={Transactions of the Association for Computational Linguistics},
  volume={7},
  pages={453--466},
  year={2019},
  publisher={MIT Press One Rogers Street, Cambridge, MA 02142-1209, USA journals-info~…},
  url={https://aclanthology.org/Q19-1026.pdf}
}

@article{jiang2023mistral,
  title={Mistral 7B},
  author={Jiang, Albert Q and Sablayrolles, Alexandre and Mensch, Arthur and Bamford, Chris and Chaplot, Devendra Singh and Casas, Diego de las and Bressand, Florian and Lengyel, Gianna and Lample, Guillaume and Saulnier, Lucile and others},
  journal={arXiv preprint arXiv:2310.06825},
  year={2023},
  url={https://arxiv.org/pdf/2310.06825}
}

@article{chatgpt,
    author = {OpenAI},
    title = {Introducing ChatGPT},
    year = {2022},
    url = {openai.com/blog/chatgpt.}
}

@article{llama3,
    author = {Meta},
    title = {Welcome Llama 3 - Meta’s new open LLM},
    year ={2024} ,
    url = {https://github.com/meta-llama/llama3}
}

@article{ren2021rocketqav2,
  title={RocketQAv2: A Joint Training Method for Dense Passage Retrieval and Passage Re-ranking},
  author={Ren, Ruiyang and Qu, Yingqi and Liu, Jing and Zhao, Wayne Xin and She, Qiaoqiao and Wu, Hua and Wang, Haifeng and Wen, Ji-Rong},
  journal={EMNLP},
  year={2021},
  url={https://aclanthology.org/2021.emnlp-main.224.pdf}
}

@article{schamber1988relevance,
  title={Relevance: The Search for a Definition.},
  author={Schamber, Linda and Eisenberg, Michael},
  year={1988},
  publisher={ERIC}
}

@article{schamber1990re,
  title={A re-examination of relevance: toward a dynamic, situational definition},
  author={Schamber, Linda and Eisenberg, Michael B and Nilan, Michael S},
  journal={Information processing \& management},
  volume={26},
  number={6},
  pages={755--776},
  year={1990},
  publisher={Elsevier},
  url={https://www.sciencedirect.com/science/article/abs/pii/030645739090050C}
}

@book{schutz2011reflections,
	address = {Westport, Conn.},
	author = {Alfred Schutz},
	editor = {Richard M. Zaner},
	publisher = {Greenwood Press},
	title = {Reflections on the Problem of Relevance},
	year = {1970}
}

@article{li2023llatrieval,
  title={Llatrieval: Llm-verified retrieval for verifiable generation},
  author={Li, Xiaonan and Zhu, Changtai and Li, Linyang and Yin, Zhangyue and Sun, Tianxiang and Qiu, Xipeng},
  journal={arXiv preprint arXiv:2311.07838},
  year={2023},
  url={https://arxiv.org/pdf/2311.07838}
}

@inproceedings{shao-etal-2023-enhancing,
    title = "Enhancing Retrieval-Augmented Large Language Models with Iterative Retrieval-Generation Synergy",
    author = "Shao, Zhihong  and
      Gong, Yeyun  and
      Shen, Yelong  and
      Huang, Minlie  and
      Duan, Nan  and
      Chen, Weizhu",
    editor = "Bouamor, Houda  and
      Pino, Juan  and
      Bali, Kalika",
    booktitle = "Findings of the Association for Computational Linguistics: EMNLP 2023",
    month = dec,
    year = "2023",
    address = "Singapore",
    publisher = "Association for Computational Linguistics",
    url = "https://aclanthology.org/2023.findings-emnlp.620",
    doi = "10.18653/v1/2023.findings-emnlp.620",
    pages = "9248--9274",
}

@inproceedings{lin2004rouge,
  title={Rouge: A package for automatic evaluation of summaries},
  author={Lin, Chin-Yew},
  booktitle={Text summarization branches out},
  pages={74--81},
  year={2004},
  url={https://aclanthology.org/W04-1013.pdf}
}

@article{robertson2009probabilistic,
  title={The probabilistic relevance framework: BM25 and beyond},
  author={Robertson, Stephen and Zaragoza, Hugo and others},
  journal={Foundations and Trends{\textregistered} in Information Retrieval},
  volume={3},
  number={4},
  pages={333--389},
  year={2009},
  publisher={Now Publishers, Inc.},
  url={https://dl.acm.org/doi/10.1561/1500000019}
}

@article{singh2023beyond,
  title={Beyond human data: Scaling self-training for problem-solving with language models},
  author={Singh, Avi and Co-Reyes, John D and Agarwal, Rishabh and Anand, Ankesh and Patil, Piyush and Liu, Peter J and Harrison, James and Lee, Jaehoon and Xu, Kelvin and Parisi, Aaron and others},
  journal={arXiv preprint arXiv:2312.06585},
  year={2023},
  url={https://arxiv.org/pdf/2312.06585}
}

@inproceedings{zhang2023relevance,
    title = "From Relevance to Utility: Evidence Retrieval with Feedback for Fact Verification",
    author = "Zhang, Hengran  and
      Zhang, Ruqing  and
      Guo, Jiafeng  and
      de Rijke, Maarten  and
      Fan, Yixing  and
      Cheng, Xueqi",
    editor = "Bouamor, Houda  and
      Pino, Juan  and
      Bali, Kalika",
    booktitle = "Findings of the Association for Computational Linguistics: EMNLP 2023",
    month = dec,
    year = "2023",
    address = "Singapore",
    publisher = "Association for Computational Linguistics",
    url = "https://aclanthology.org/2023.findings-emnlp.422",
    doi = "10.18653/v1/2023.findings-emnlp.422",
    pages = "6373--6384",
}

@article{mizzaro1997relevance,
  title={Relevance: The whole history},
  author={Mizzaro, Stefano},
  journal={Journal of the American society for information science},
  volume={48},
  number={9},
  pages={810--832},
  year={1997},
  publisher={Wiley Online Library}
}

@article{bruce1994cognitive,
  title={A cognitive view of the situational dynamism of user-centered relevance estimation},
  author={Bruce, Harry W},
  journal={Journal of the American Society for Information Science},
  volume={45},
  number={3},
  pages={142--148},
  year={1994},
  publisher={Wiley Online Library}
}

@inproceedings{vickery1959structure,
  title={The structure of information retrieval systems},
  author={Vickery, Brian C},
  booktitle={Proceedings of the International Conference on Scientific Information},
  volume={2},
  pages={1275--1290},
  year={1959}
}

@article{hillman1964notion,
  title={The notion of relevance (I)},
  author={Hillman, Donald J},
  journal={American Documentation},
  volume={15},
  number={1},
  pages={26--34},
  year={1964},
  publisher={Wiley Online Library}
}

@article{cooper1971definition,
  title={A definition of relevance for information retrieval},
  author={Cooper, William S},
  journal={Information storage and retrieval},
  volume={7},
  number={1},
  pages={19--37},
  year={1971},
  publisher={Elsevier}
}

@article{swanson1986subjective,
  title={Subjective versus objective relevance in bibliographic retrieval systems},
  author={Swanson, Don R},
  journal={The library quarterly},
  volume={56},
  number={4},
  pages={389--398},
  year={1986},
  publisher={University of Chicago Press}
}

@article{lancaster1968information,
  title={Information retrieval systems: Characteristics, testing, and evaluation},
  author={Lancaster, F Wilfrid},
  journal={(No Title)},
  year={1968}
}

@book{goffman1964methodology,
  title={Methodology for test and evaluation of information retrieval systems},
  author={Goffman, William and Newill, Vaun A},
  year={1964},
  publisher={Center for Documentation and Communication Research, School of Library~…}
}

@article{kemp1974relevance,
  title={Relevance, pertinence and information system development},
  author={Kemp, DA},
  journal={Information Storage and Retrieval},
  volume={10},
  number={2},
  pages={37--47},
  year={1974},
  publisher={Elsevier}
}

@article{mizzaro1998many,
  title={How many relevances in information retrieval?},
  author={Mizzaro, Stefano},
  journal={Interacting with computers},
  volume={10},
  number={3},
  pages={303--320},
  year={1998},
  publisher={OUP}
}

@article{yang2025qwen3,
  title={Qwen3 technical report},
  author={Yang, An and Li, Anfeng and Yang, Baosong and Zhang, Beichen and Hui, Binyuan and Zheng, Bo and Yu, Bowen and Gao, Chang and Huang, Chengen and Lv, Chenxu and others},
  journal={arXiv preprint arXiv:2505.09388},
  year={2025}
}

@article{wang2022self,
  title={Self-consistency improves chain of thought reasoning in language models},
  author={Wang, Xuezhi and Wei, Jason and Schuurmans, Dale and Le, Quoc and Chi, Ed and Narang, Sharan and Chowdhery, Aakanksha and Zhou, Denny},
  journal={arXiv preprint arXiv:2203.11171},
  year={2022}
}
\appendix
\clearpage
\begin{table}[t]
  \centering
   \renewcommand{\arraystretch}{0.95}
   \setlength\tabcolsep{2pt}
    \begin{tabular}{ccccc}
    \toprule
    Dataset & \#Psg  & \#PsgLen  & \#Q  & \#Rels/Q  \\
    \midrule
    TREC  & 8.8M  & 93    & 82    & 212.8 \\
    WebAP & 379k  & 45    & 73    & 76.4 \\
    NQ    & 21M   & 100   & 1868  & 1.0 \\
    GTI-NQ & 10  & 100 & 1863 & 1.0 \\
    \bottomrule
    \end{tabular}%
    \caption{Statistics of experimental datasets.}
  \label{app:tab:dataset}%
\end{table}%
\section{Datasets and Evaluation}
\label{app:dataset}
Detailed statistics of the experimental datasets are shown in Table \ref{app:tab:dataset}. 
We use the $trec\_eval$ official tool for evaluation of NDCG. 
The annotation instruction for different points from \citet{craswell2020overview}: 
\begin{itemize}
    \item  Perfectly relevant: The passage is dedicated to the query and contains the exact answer. 
    \item Highly relevant: The passage has some answer for the query, but the answer may be a bit unclear, or hidden amongst extraneous information.
    \item Related: The passage seems related to the query but does not answer it. 
    \item Irrelevant: The passage has nothing to do with the query.
\end{itemize}

The highest level of relevance can be seen as aligning with the definition of utility; therefore, we use passages labeled with the highest relevance level in TREC DL and WebAP as positive examples for utility judgments evaluation.

\section{Experiment Details}
\subsection{Effect of Iteration Numbers}
\label{app:exp:m_values}
The precision, recall, and F1 performance of different LLMs on different datasets with different iteration numbers is shown in Table \ref{tab:app:msitral}, Table \ref{tab:app:llama}, Table \ref{tab:app:chatgpt}, and Table  \ref{tab:app:nq}. 
\begin{table*}[htbp]
  \centering
  \small
    \renewcommand{\arraystretch}{0.95}
   \setlength\tabcolsep{2pt}
    \begin{tabular}{c ccc ccc ccc ccc}
    \toprule
    \multicolumn{1}{c}{\multirow{3}[6]{*}{Method}} & \multicolumn{6}{c}{TREC}                     & \multicolumn{6}{c}{WebAP} \\
\cmidrule(r){2-7}   \cmidrule(r){8-13}  \multicolumn{1}{c}{} & \multicolumn{3}{c}{listwise} & \multicolumn{3}{c}{pointwise} & \multicolumn{3}{c}{listwise} & \multicolumn{3}{c}{pointwise} \\
\cmidrule(r){2-4} \cmidrule(r){5-7} \cmidrule(r){8-10} \cmidrule(r){11-13}    \multicolumn{1}{c}{} & P     & R     & F1    & P     & R     & F1    & P     & R     & F1    & P     & R     & F1 \\
    \midrule
     Vanilla  & 36.82  & 60.13  & 45.67  & 29.92  & \textbf{91.61}  & 45.11  & 13.07  & 50.83  & 20.79  & 13.30  & 86.29  &  23.05 \\
    \midrule
     UJ-ExpA & 48.51  & 61.15  & 54.10  & 28.12  & 96.27  & 43.53  & 18.83  & 54.16  & 27.94  & 14.65  & \textbf{91.82}  & 25.27  \\
     UJ-ImpA  & 40.16  & 60.53  & 48.29  & 33.95  & 83.76  & 48.31  & 16.46  & 52.45  & 25.06  & 17.56  & 73.55  & 28.35  \\
      5-sampling  & 46.64 	 &59.56  &52.31  &     -    &  -     &    - 	 &	20.61 &56.22  &	30.16   &     -    &  -     &    -    \\
    \midrule
    \modelname-A$_s$ $w.$ ExpA ($m$=1) & 48.07  & 61.04  & 53.78  & 34.21  & 89.11  & 49.44  & 20.57  & 53.81  & 29.76  & 17.86  & 78.41  & 29.10  \\
    \modelname-A$_s$ $w.$ ExpA ($m$=2) & 50.58  & 61.86  & 55.65  & 35.87  & 88.73  & 51.09  & 21.11  & 50.85  & 29.83  & 18.27  & 82.00  & 29.88  \\
    \modelname-A$_s$ $w.$ ExpA ($m$=3) & 50.61  & 59.88  & 54.86  & 36.23  & 90.46  & 51.74  & 23.57  & 48.14  & 31.65  & 18.73  & 81.96  & 30.50  \\
    \modelname-A$_s$ $w.$ ExpA ($m$=4) & 50.01  & 61.15  & 55.02  & \textbf{36.41}  & 90.36  & \textbf{51.90}  & 21.44  & 44.62  & 28.96  & \textbf{19.19}  & 80.59  & \textbf{31.00}  \\
    \modelname-A$_s$ $w.$ ExpA ($m$=5) & 50.61  & 59.88  & 54.86  & 36.14  & 90.46  & 51.65  & 24.07  & 47.09  & 31.86  & 19.17  & 78.94  & 30.85  \\
    \midrule
    \modelname-A$_s$ $w.$ ImpA ($m$=1) & 39.97  & 64.62  & 49.39  & 30.98  & 89.38  & 46.01  & 16.88  & 57.13  & 26.06  & 17.10  & 81.65  & 28.28  \\
    \modelname-A$_s$ $w.$ ImpA ($m$=2) & 43.14  & 61.52  & 50.72  & 30.90  & 87.00  & 45.60  & 19.41  & 54.82  & 28.67  & 18.88  & 78.06  & 30.40  \\
    \modelname-A$_s$ $w.$ ImpA ($m$=3) & 44.43  & 62.82  & 52.05  & 31.68  & 87.99  & 46.59  & 19.21  & 54.20  & 28.36  & 18.69  & 77.77  & 30.13  \\
    \modelname-A$_s$ $w.$ ImpA ($m$=4) & 44.72  & 61.29  & 51.71  & 31.66  & 87.40  & 46.49  & 17.44  & 47.11  & 25.46  & 18.95  & 78.06  & 30.50  \\
    \modelname-A$_s$ $w.$ ImpA ($m$=5) & 44.63  & 60.98  & 51.54  & 31.80  & 89.32  & 46.91  & 18.98  & 48.88  & 27.35  & 19.05  & 76.69  & 30.52  \\
    \midrule
    \modelname-AR$_s$ ($m$=1) & 43.65  & 65.34  & 52.34  &     -    &  -     &    -     & 25.04  & \textbf{60.99}  & 35.50  &    -    &  -     &    -    \\
    \modelname-AR$_s$ ($m$=2) & 45.10  & 65.46  & 53.40  &      -    &  -     &    -     & 24.42  & 51.97  & 33.23  &     -    &  -     &    -    \\
    \modelname-AR$_s$  ($m$=3) & 49.07  & \textbf{65.96}  & 56.27  &      -    &  -     &    -    & \textbf{27.70}  & 55.95  & \textbf{37.06}  &  -    &  -     &    -    \\
    \modelname-AR$_s$  ($m$=4) & 50.96  & 62.32  & 56.07  &    -    &  -     &    -        & 23.77  & 53.40  & 32.90  &       -    &  -     &    -   \\
    \modelname-AR$_s$  ($m$=5) & \textbf{53.01}  & 63.60  & \textbf{57.82}  &     -    &  -     &    -     & 25.85  & 47.56  & 33.50  &     -    &  -     &    -   \\
    \bottomrule
    \end{tabular}%
    \caption{The utility judgments performance (\%) of Mistral on retrieval datasets (Numbers in parentheses represent $m$-values). Numbers in bold indicate the best performance.}
  \label{tab:app:msitral}%
\end{table*}%

\begin{table*}[htbp]
  \centering
  \small
    \renewcommand{\arraystretch}{0.95}
   \setlength\tabcolsep{2pt}
    \begin{tabular}{c ccc ccc ccc ccc}
    \toprule
       \multicolumn{1}{c}{\multirow{3}[6]{*}{Method}} & \multicolumn{6}{c}{TREC}                     & \multicolumn{6}{c}{WebAP} \\
\cmidrule(r){2-7}   \cmidrule(r){8-13}  \multicolumn{1}{c}{} & \multicolumn{3}{c}{listwise} & \multicolumn{3}{c}{pointwise} & \multicolumn{3}{c}{listwise} & \multicolumn{3}{c}{pointwise} \\
\cmidrule(r){2-4} \cmidrule(r){5-7} \cmidrule(r){8-10} \cmidrule(r){11-13}    \multicolumn{1}{c}{} & P     & R     & F1    & P     & R     & F1    & P     & R     & F1    & P     & R     & F1 \\
    \midrule
     Vanilla & 34.67  & \textbf{85.80}  & 49.39  & 31.42  & \textbf{98.47}  & 47.64  & 12.69  & 77.15  & 21.79  & 14.65  & \textbf{87.36}  & 25.09 \\
    \midrule
   UJ-ExpA & 39.21  & 80.98  & 52.83  & 38.27  & 90.15  & 53.73  & 16.32  & 77.92  & 26.99  & 18.04  & 77.15  & 29.25  \\
     UJ-ImpA  & 33.92  & 83.36  & 48.22  & 38.68  & 71.47  & 50.20  & 15.57  & \textbf{82.79}  & 26.22  & 17.22  & 47.61  & 25.29  \\
   5-sampling  &  39.04  &	80.98  &	52.68 & - & - & -				& 17.52 	& 83.49 &	28.97 & - & - & - \\
    \midrule
    \modelname-A$_s$ $w.$ ExpA ($m$=1) & 39.68  & 82.88  & 53.66  & 37.58  & 84.84  & 52.09  & 17.54  & 63.67  & 27.50  & 19.65  & 74.31  & 31.08  \\
     \modelname-A$_s$ $w.$ ExpA ($m$=2)  & \textbf{42.35}  & 84.77  & \textbf{56.48}  & 38.25  & 84.58  & 52.68  & 17.39  & 60.25  & 26.99  & 20.23  & 73.01  & 31.68  \\
     \modelname-A$_s$ $w.$ ExpA ($m$=3)  & 42.00  & 84.15  & 56.03  & 37.84  & 85.50  & 52.46  & 19.12  & 62.87  & 29.32  & \textbf{20.91}  & 74.63  & 32.67  \\
     \modelname-A$_s$ $w.$ ExpA ($m$=4)  & 41.85  & 84.41  & 55.96  & 38.12  & 85.16  & 52.67  & 17.53  & 61.85  & 27.31  & 20.44  & 73.83  & 32.02  \\
     \modelname-A$_s$ $w.$ ExpA ($m$=5)  & 42.36  & 84.15  & 56.35  & 37.35  & 84.69  & 51.84  & 18.94  & 62.87  & 29.12  & 20.88  & 75.45  & \textbf{32.71}  \\
    \midrule
    \modelname-A$_s$ $w.$ ImpA ($m$=1) & 39.63  & 83.42  & 53.73  & 39.70  & 82.87  & 53.68  & 15.48  & 73.66  & 25.59  & 20.04  & 64.06  & 30.53  \\
    \modelname-A$_s$ $w.$ ImpA ($m$=2) & 38.75  & 85.63  & 53.35  & 38.15  & 82.36  & 52.14  & 15.50  & 76.47  & 25.77  & 18.54  & 62.69  & 28.62  \\
    \modelname-A$_s$ $w.$ ImpA ($m$=3) & 40.84  & 84.86  & 55.14  & 40.58  & 79.64  & 53.76  & 15.99  & 70.99  & 26.10  & 19.54  & 61.32  & 29.64  \\
    \modelname-A$_s$ $w.$ ImpA ($m$=4) & 38.88  & 82.74  & 52.90  & 39.34  & 81.74  & 53.12  & 15.03  & 74.41  & 25.01  & 19.72  & 59.95  & 29.68  \\
    \modelname-A$_s$ $w.$ ImpA ($m$=5) & 41.26  & 84.61  & 55.47  & \textbf{40.92}  & 82.14  & \textbf{54.63}  & 15.49  & 68.93  & 25.29  & 19.84  & 57.21  & 29.46  \\
    \midrule
    \modelname-AR$_s$ ($m$=1) & 34.53  & 84.17  & 48.97  &      -    &  -     &    -       & \textbf{20.05}  & 72.88  & \textbf{31.44}  &   -    &  -     &    -   \\
    \modelname-AR$_s$  ($m$=2)& 36.27  & 83.19  & 50.51   	    &   -    &  -     &    -       & 15.92  & 79.01  & 26.50  &       -    &  -     &    -   \\
    \modelname-AR$_s$   ($m$=3) & 38.04  & 82.68  & 52.10 &  -    &  -     &    -       & 17.93  & 76.87  & 29.08  &       -    &  -     &    -   \\
    \modelname-AR$_s$   ($m$=4) & 37.28  & 83.70  & 51.58    &   -    &  -     &    -      & 16.60  & 78.81  & 27.42  &      -    &  -     &    -   \\
    \modelname-AR$_s$   ($m$=5) & 40.25  & 81.37  & 53.86 &     -    &  -     &    -     & 17.04  & 74.83  & 27.75  &      -    &  -     &    -   \\
    \bottomrule
    \end{tabular}%
    \caption{The utility judgments performance (\%) of Llama 3 on retrieval datasets (Numbers in parentheses represent $m$-values). Numbers in bold indicate the best performance.}
  \label{tab:app:llama}%
\end{table*}%

\begin{table*}[htbp]
  \centering
  \small
    \renewcommand{\arraystretch}{0.95}
   \setlength\tabcolsep{2pt}
    \begin{tabular}{cccc ccc ccc ccc}
    \toprule
        \multicolumn{1}{c}{\multirow{3}[6]{*}{Method}} & \multicolumn{6}{c}{TREC}                     & \multicolumn{6}{c}{WebAP} \\
\cmidrule(r){2-7}   \cmidrule(r){8-13}  \multicolumn{1}{c}{} & \multicolumn{3}{c}{listwise} & \multicolumn{3}{c}{pointwise} & \multicolumn{3}{c}{listwise} & \multicolumn{3}{c}{pointwise} \\
\cmidrule(r){2-4} \cmidrule(r){5-7} \cmidrule(r){8-10} \cmidrule(r){11-13}    \multicolumn{1}{c}{} & P     & R     & F1    & P     & R     & F1    & P     & R     & F1    & P     & R     & F1 \\
    \midrule
    Vanilla & 42.13  & \textbf{79.98}  & 55.19  & 33.86  & 94.40  & 49.84  & 17.13  & \textbf{83.45}  & 28.43  & 15.80  & \textbf{89.42}  & 26.85  \\
    \midrule
   UJ-ExpA & 45.74  & 77.36  & 57.49  & 32.06  & \textbf{96.19}  & 48.09  & 19.51  & 69.86  & 30.50  & 16.23  & 88.74  & 27.44  \\
    UJ-ImpA & 44.19  & 77.11  & 56.18  & 33.45  & 90.36  & 48.83  & 18.37  & 80.14  & 29.89  & 15.58  & 84.51  & 26.32  \\
   5-sampling & 50.78 &	74.77 &	60.49 	& - & - & - &			20.70 &	65.83 &	31.49 & - & - & - \\
    \midrule
    \modelname-A$_s$ $w.$ ExpA ($m$=1) & 55.55  & 71.48  & 62.52  & 37.83  & 91.94  & 53.61  & 26.74  & 59.45  & 36.89  & 19.73  & 84.95  & 32.02  \\
    \modelname-A$_s$ $w.$ ExpA ($m$=2) & 57.95  & 70.40  & 63.57  & 40.74  & 93.04  & 56.67  & 29.43  & 60.58  & 39.62  & 19.62  & 78.62  & 31.40  \\
    \modelname-A$_s$ $w.$ ExpA ($m$=3) & 58.36  & 68.88  & 63.18  & 40.00  & 91.88  & 55.74  & 29.30  & 60.91  & 39.57  & 19.80  & 76.20  & 31.43  \\
    \modelname-A$_s$ $w.$ ExpA ($m$=4) & \textbf{58.48}  & 70.67  & \textbf{64.00}  & 40.25  & 93.38  & 56.25  & 29.11  & 61.03  & 39.42  & 20.48  & 79.63  & 32.58  \\
    \modelname-A$_s$ $w.$ ExpA ($m$=5) & 58.34  & 69.69  & 63.51  & 39.29  & 92.16  & 55.09  & 29.76  & 60.68  & \textbf{39.93}  & 20.58  & 80.42  & \textbf{32.77}  \\
    \midrule
    \modelname-A$_s$ $w.$ ImpA ($m$=1) & 54.36  & 65.08  & 59.24  & 40.89  & 82.20  & 54.61  & 24.79  & 64.37  & 35.80  & 18.78  & 67.00  & 29.34  \\
    \modelname-A$_s$ $w.$ ImpA ($m$=2) & 55.88  & 63.11  & 59.27  & 43.32  & 83.13  & \textbf{56.96}  & 27.68  & 62.03  & 38.28  & 20.70  & 70.54  & 32.00  \\
    \modelname-A$_s$ $w.$ ImpA ($m$=3) & 57.33  & 64.17  & 60.56  & 41.66  & 80.48  & 54.90  & \textbf{30.01}  & 63.60  & 40.78  & 21.51  & 66.77  & 32.54  \\
   \modelname-A$_s$ $w.$ ImpA ($m$=4) & 55.98  & 62.24  & 58.95  & \textbf{42.34}  & 80.65  & 55.53  & 28.43  & 60.11  & 38.60  & 20.60  & 65.63  & 31.36  \\
    \modelname-A$_s$ $w.$ ImpA ($m$=5) & 56.63  & 62.19  & 59.28  & 41.49  & 83.57  & 55.45  & 29.05  & 60.66  & 39.29  & \textbf{21.51}  & 68.03  & 32.68  \\
    \midrule
    \modelname-AR$_s$ ($m$=1) & 51.94  & 76.90  & 62.00  &   -    &  -     &    -   & 25.32  & 65.84  & 36.58  &     -    &  -     &    -     \\
    \modelname-AR$_s$  ($m$=2) & 53.77  & 76.19  & 63.05  &   -    &  -     &    -    & 25.55  & 59.26  & 35.70  &   -    &  -     &    -     \\
    \modelname-AR$_s$   ($m$=3) & 52.41  & 74.04  & 61.37  &   -    &  -     &    -   & 27.61  & 63.96  & 38.58  &   -    &  -     &    -   \\
    \modelname-AR$_s$   ($m$=4) & 52.75  & 73.78  & 61.52  &     -    &  -     &    -      & 28.84  & 61.85  & 39.34  &     -    &  -     &    -   \\
    \modelname-AR$_s$   ($m$=5) & 52.77  & 76.28  & 62.39  &    -    &  -     &    -      & 28.76  & 62.54  & 39.40  &     -    &  -     &    -   \\
    \bottomrule
    \end{tabular}%
    \caption{The utility judgments performance of ChatGPT on retrieval datasets (Numbers in parentheses represent $m$-values). Numbers in bold indicate the best performance.}
  \label{tab:app:chatgpt}%
\end{table*}%

\begin{table*}[htbp]
  \centering
    \begin{tabular}{ccccccc}
    \toprule
   \multirow{2}[3]{*}{References of RAG} & \multicolumn{2}{c}{Mistral} & \multicolumn{2}{c}{Llama 3} & \multicolumn{2}{c}{ChatGPT} \\
\cmidrule(r){2-3}   \cmidrule(r){4-5}  \cmidrule(r){6-7}       & EM    & F1    & EM    & F1 & EM    & F1\\
\midrule
    Golden Evidence & 46.09  & 62.59  & 64.45  & 76.64  & 66.40  & 76.86  \\
    RocketQAv2 & 31.58  & 47.69  & \textbf{50.96 } & 62.01  & 46.54  & 57.00  \\
    \midrule
    Vanilla & 31.16  & 47.43  & 49.09  & 60.56  & 48.52  & 58.64  \\
    UJ-ExpA & 32.76  & 48.46  & 49.63  & 61.10  & 47.72  & 58.01  \\
    UJ-ImpA & 30.67  & 46.83  & 48.88  & 60.26  & 49.01  & 59.30  \\
    5-sampling & 33.24  & 48.84  & 48.72  & 60.71  & 48.90  & 58.97  \\
    \midrule
    \modelname-A$_s$ $w.$ ExpA ($m$=1) & 32.98  & 49.00  & 50.16  & 61.88  & 49.38 & 59.78 \\
    \modelname-A$_s$ $w.$ ExpA ($m$=2) & \textbf{34.31} & \textbf{50.08} & 50.48  & \textbf{62.32} & 49.22  & 59.99  \\
    \modelname-A$_s$ $w.$ ExpA ($m$=3) & 33.73  & 49.63  & 50.27  & 62.09  & \textbf{49.69} & \textbf{60.18} \\
    \modelname-A$_s$ $w.$ ExpA ($m$=4) & 34.21  & 50.07  & 50.43  & 62.20  &    -   &  -\\
    \modelname-A$_s$ $w.$ ExpA ($m$=5) & 33.78  & 49.63  & 50.27  & 62.07  &  -     & - \\
    \midrule
    \modelname-A$_s$ $w.$ ImpA ($m$=1) & 32.17  & 48.51  & 50.37  & 61.89  & 48.75 & 58.99 \\
    \modelname-A$_s$ $w.$ ImpA ($m$=2) & 32.49  & 48.67 & 49.63  & 61.16  & 49.11 & 59.14 \\
    \modelname-A$_s$ $w.$ ImpA ($m$=3) & 32.39  & 48.47  & 49.68  & 61.48  & 48.69 & 58.94 \\
    \modelname-A$_s$ $w.$ ImpA ($m$=4) & 32.71  & 48.84  & 49.41  & 61.03  &   -   & - \\
    \modelname-A$_s$ $w.$ ImpA ($m$=5) & 32.33  & 48.44  & 49.73  & 61.42  &   -    & - \\
    \midrule
    \modelname-AR$_s$ ($m$=1) & 33.30  & 49.26  & 50.27  & 61.69  & 49.52 & 59.64 \\
    \modelname-AR$_s$ ($m$=2) & 33.57  & 49.16  & 50.70  & 61.92  & 49.01 & 59.75 \\
    \modelname-AR$_s$ ($m$=3) & 33.40  & 49.27  & 49.36  & 60.97  & 49.06 & 59.67 \\
    \modelname-AR$_s$ ($m$=4) & 33.46  & 49.24  & 49.84  & 61.54  &     -  & - \\
    \modelname-AR$_s$ ($m$=5) & 33.89  & 49.58  & 49.20  & 60.84  &     -  & - \\
    \bottomrule
    \end{tabular}%
   \caption{The answer generation performance (\%) of all LLMs in the listwise approach. Numbers in bold indicate the best performance except the answer performance using golden evidence. Due to the high cost of using ChatGPT, we only tested with $m$=1,2,3 on ChatGPT. }
  \label{tab:app:nq}%
\end{table*}%

\subsection{Different Multiple Calls methods}
\begin{table}[t]
\centering
\small
\setlength\tabcolsep{3pt}
\caption{F1 Performance (\%) comparison of different methods on TREC DL and WebAP datasets using listwise approach upon Llama-3.1-8B($T$ is temperature during generation, $k$ is top-$k$ sampling, $p$ is nucleus sampling). }
\label{tab:results}
\begin{tabular}{lll}
\toprule
\textbf{Method} & \textbf{TREC DL} & \textbf{WebAP} \\
\midrule
UJ-ExpA (T=0) & 52.83 & 26.99 \\
$k$-sampling (T=0) & 52.68 & 28.97 \\
\midrule
SC ($T=0.7$, $k=40$) & 54.17 & 26.11 \\
SC ($T=0.5$, $k=40$) & 52.71 & 27.08 \\
SC ($T=0.3$, $k=40$) & 53.76 & 25.79 \\
SC ($T=0.7$, $k=20$) & 54.39 & 26.24 \\
SC ($T=0.7$, $k=40$, $p=0.95$) & 53.31 & 27.08 \\
SC ($T=0.7$, $k=40$, $p=0.90$) & 54.32 & 26.67 \\
\midrule
\modelname-A$_s$ (T=0, $m=3$) & \textbf{56.03} & \textbf{29.32} \\
\bottomrule
\end{tabular}
\label{app:more_baselines}
\end{table}
For methods that involve multiple calls for better utility judgment performance, $k$-sampling (our baseline) randomly shuffles the input and aggregates multiple results to obtain the final utility judgments. 
We also reproduced another multiple calls method, the self-consistency method (SC) \cite{wang2022self},  on the utility judgments task. 
That is, we sampled multiple generations from the same input (The prompt is the same as UJ-ExpA) and aggregated them to get the final utility judgments. Following the $k$-sampling method, we used 5 utility judgment results and aggregated them in a way similar to $k$-sampling to obtain the final utility judgments. The results are shown in Table \ref{app:more_baselines}. We can observe that our \modelname achieves the best performance, indicating the superior performance of iterative judgments. 

\subsection{$k$ values in \modelname-A$_r$}
\label{app:k_in_item-ar}
Different ranking performance of $k$ values in \modelname-A$_r$ is shown in Table \ref{tab:app:p_values}. 
Considering the performance of utility ranking and utility judgments, we set $k$=5.
\label{app:p_values}
\begin{table*}[htbp]
  \centering
    \begin{tabular}{c ccccc ccc}
    \toprule
    \multirow{2}[4]{*}{$k$, $m$} & \multicolumn{5}{c}{Ranking} & \multicolumn{3}{c}{Utility judgments} \\
\cmidrule(r){2-6}  \cmidrule(r){7-9}  & N@1  & N@3  & N@5  & N@10 & N@20 & P     & R     & F1 \\
    \midrule
    $k$=1, $m$=1 & 72.76  & 71.27  & 70.57  & 72.69  & 84.08  & 53.66  & 24.09  & 33.25  \\
    $k$=1, $m$=2 & 76.02  & 71.54  & 71.38  & 73.66  & 84.78  & 58.54  & 28.73  & 38.54  \\
    $k$=1, $m$=3 & 77.24  & 72.83  & 71.83  & 73.87  & 85.20  & \textbf{59.76} & 28.84  & 38.90  \\
    $k$=1, $m$=4 & 77.24  & 73.04  & 71.91  & 73.90  & 85.25  & 59.76  & 28.84  & 38.90  \\
    $k$=1, $m$=5 & 76.02  & 72.11  & 71.42  & 73.45  & 84.98  & 58.54  & 28.71  & 38.53  \\
    \midrule
    $k$=5, $m$=1 & 72.76  & 71.27  & 70.57  & 72.69  & 84.08  & 33.17  & 57.31  & 42.02  \\
    $k$=5, $m$=2 & 78.46  & 73.74  & 72.86  & 75.48  & 86.09  & 32.93  & 58.37  & 42.10  \\
    $k$=5, $m$=3 & 79.27  & 75.00  & 74.27  & 75.78  & 86.80  & 34.15  & 62.57  & 44.18  \\
    $k$=5, $m$=4 & 79.67  & 75.92  & \textbf{75.35} & \textbf{76.83} & \textbf{87.23} & 35.12  & 61.40  & \textbf{44.68} \\
   $k$=5, $m$=5 & 79.67  & 75.32  & 74.61  & 76.20  & 86.82  & 34.63  & 61.25  & 44.25  \\
    \midrule
    $k$=10, $m$=1 & 72.76  & 71.27  & 70.57  & 72.69  & 84.08  & 22.56  &  68.03 & 33.88  \\
    $k$=10, $m$=2 & 78.05  & 72.64  & 72.90  & 75.48  & 85.74  & 23.66  & 75.47  & 36.02  \\
   $k$=10, $m$=3 & \textbf{80.89} & \textbf{76.58} & 74.54  & 76.30  & 86.94  & 23.78  & \textbf{75.65}  & 36.19  \\
    $k$=10, $m$=4 & 78.05  & 74.70  & 72.85  & 75.12  & 85.72  & 24.51  & 74.17  & 36.85  \\
    $k$=10, $m$=5 & 79.67  & 75.60  & 74.84  & 76.54  & 86.88  & 23.66  & 74.42  & 35.90  \\
    \bottomrule
    \end{tabular}%
    \caption{The utility ranking performance and utility judgments performance of Mistral on \trec dataset in \modelname-A$_r$. ``N@k'' means ``NDCG@k''. Numbers in bold indicate the best performance.}
  \label{tab:app:p_values}%
\end{table*}%

\subsection{\modelname-AR$_r$}
\label{app:item-arr}
We evaluate two ranking performances of \modelname-AR$_r$ during the same loop, with the experimental results shown in Table \ref{tab:iteration_rudif}. 
We find that under the \modelname-AR$_r$ framework, relevance ranking and utility ranking are both improved, and utility ranking performance is generally better than relevance ranking. 
However, as seen in Table \ref{tab:ranking_msitral} and Table \ref{tab:iteration_rudif}, performing ranking twice in the same iteration may not yield better ranking results than performing utility ranking once in the iteration.

\begin{table}[t]
     \small
  \centering
   \setlength\tabcolsep{3pt}
    \begin{tabular}{cccc c}
    \toprule
    $m$ & NDCG@5 & NDCG@10 & NDCG@20 & Utility-F1\\
    \midrule
    1   & 71.29 / \underline{72.77} & 72.90 / \underline{74.96} & 84.56 / \underline{85.75} & 43.13 \\
    2   & \underline{\textbf{72.54}} / 70.99 & \underline{74.81} / 73.76 & \underline{85.77} / 85.28 & 40.21 \\
    3   & 72.07 / \underline{\textbf{74.14}} & 74.14 / \underline{\textbf{76.63}} & 85.53 / \underline{\textbf{86.57}} & \textbf{45.67} \\
    4   & 71.02 / \underline{71.06} & \underline{74.30} / 74.03 & 85.09 / \underline{85.16} & 43.82\\
    5   & \underline{72.26} / 70.12 & \underline{\textbf{75.83}} / 72.59 & \underline{\textbf{85.88}} / 84.77 & 44.10 \\
    \bottomrule
    \end{tabular}%
    \caption{Ranking of topical relevance  and utility judgments performance (\%) of \modelname-AR$_r$ using Mistral on the \trec dataset. ``a/b'' means ``relevance ranking performance / utility ranking performance''. Numbers with \underline{underline} mean better performance among all variants of \modelname with the same $m$.}
  \label{tab:iteration_rudif}%
\end{table}%

\section{Case Study}
\label{app:case_study}
\begin{figure*}[t]
    \centering
    \includegraphics[width=\linewidth]{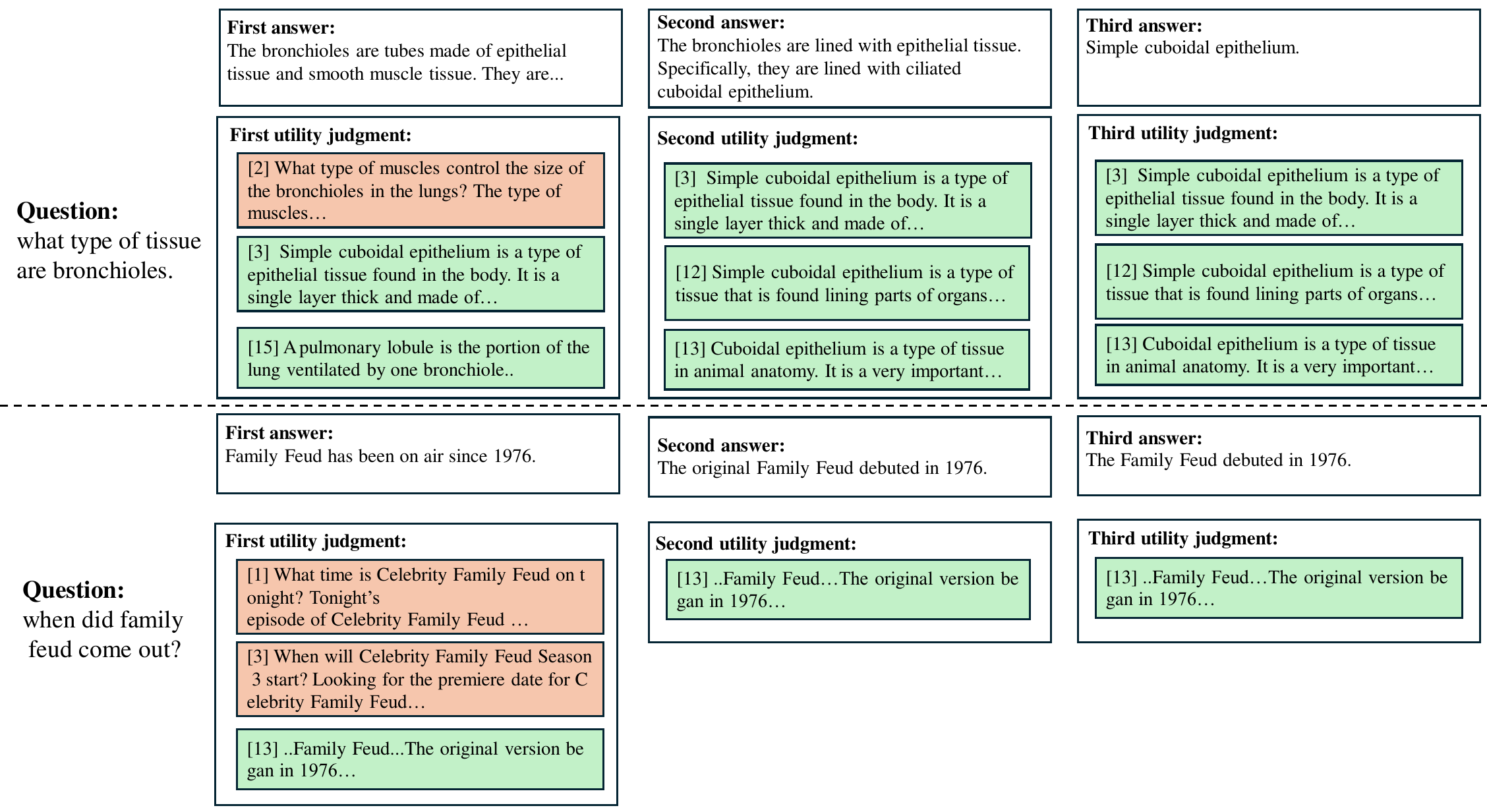}
    \caption{An example of our \modelname-A$_s$ using Mistral on the  TREC dataset. Green means the passage has utility, and orange means the passage does not have utility.}
    \label{fig:exp-case-1}
\end{figure*}

Figure \ref{fig:exp-case-1} presents two cases from the \trec dataset using Mistral under \modelname-A$_s$. 
For the first question in Figure \ref{fig:exp-case-1}, the first pseudo-answer, though relatively correct, includes irrelevant information, leading to a misjudgment of ``Passage-2'' as ``utility''. 
Based on the results of the first round of utility judgments, the second round of the pseudo-answer is more accurate and free of irrelevant content. 
Consequently, all three passages in the second round of utility judgments have utility in answering the question. 
For the second question in Figure \ref{fig:exp-case-1}, the first pseudo-answer is correct, but two passages without utility are judged as ``utility''. 
The second pseudo-answer, with slight rewording, results in all selected passages being correct.

\section{Answer Passage Retrieval}
\label{app:answer_passage_retrieval}
Non-factoid questions usually expected longer answers, such as sentence-level or passages-level \cite{keikha2014evaluating, yang2016beyond, keikha2014retrieving}.
\citet{yang2016beyond} developed an annotated dataset for answer passage retrieval called WebAP, which has an average of 76.4 qrels per query.
\citet{cohen2018wikipassageqa} and \citet{hashemi2020antique} introduced the WikiPassageQA dataset and ANTIQUE dataset for answer passage retrieval, respectively.
Compared to the WebAP dataset, WikiPassageQA and ANTIQUE have incomplete annotations, with an average of 1.7 qrels and 32.9 qrels per query \cite{hashemi2019performance, hashemi2020antique}.
\citet{bi2019iterative} created the PsgRobust dataset for answer passage retrieval, which is built on the TREC Robust collection \cite{voorhees2003overview} following a similar approach to WebAP but without manual annotation.

\section{$k$-sampling}
\label{app:k-sampling}
The output of $k$-sampling each time contains explicit answers and utility judgments. 
If the question length is $l_q$, the total length of the input passages is $l_p$, and the average length of a single passage is $l_{\text{avg}}$, then the k-sampling input cost is $(k+1) \times (l_q + l_p)$. If the average length of the output explicit answer is $l_a$, and the average length of the output utility judgments is $l_u$, then the k-sampling output cost is $(k+1) \times (l_a + l_u)$. 
Taking ITEM-As as an example, with a maximum of three iterations, the maximum input cost for utility judgments is $3 \times (l_q + l_p)$. For answer generation, the longest input is $l_q + l_p$ and the shortest is $l_q + l_{\text{avg}}$. Therefore, the maximum input cost for ITEM-As is $6 \times (l_q + l_p)$ and the minimum is $4 \times (l_q + l_p) + 2 \times (l_q + l_{\text{avg}})$. The maximum output cost is $3 \times (l_a + l_u)$. 
Therefore, in order to ensure fairness in the calculation of large language model parameters, we choose $k$=5.



\section{Retrievers}
\label{app:retrievers}
We use two representative retrievers to gather candidate passages in $\mathcal{D}$ for utility judgments.
Following with previous works \cite{Zhang2024AreLL, sun2023chatgpt}, we employ RocketQAv2 \cite{ren2021rocketqav2} and BM25 \cite{robertson2009probabilistic} for the \nq dataset and retrieval datasets(i.e., \trec and \webap datasets), respectively. 
Based on the retrieval results to build the $\mathcal{D}$ we have two settings:
\begin{enumerate*}[label=(\roman*)]
    \item For \trec and \webap datasets, we select the top-20 BM25 retrieval results. If these do not include passages with utility, we replaced the last one with a utility-annotated passage. 
    \item For the NQ dataset, we use the top-10 dense retrieval results to form the candidate list $\mathcal{D}$, following the GTU setting of \citet{Zhang2024AreLL}.
\end{enumerate*}

\section{Instruction Details}
\label{app:instruction}
\subsection{Instruction of Listwise and Pointwise Approaches}
\begin{figure*}[t]
    \centering
    \includegraphics[width=\linewidth]{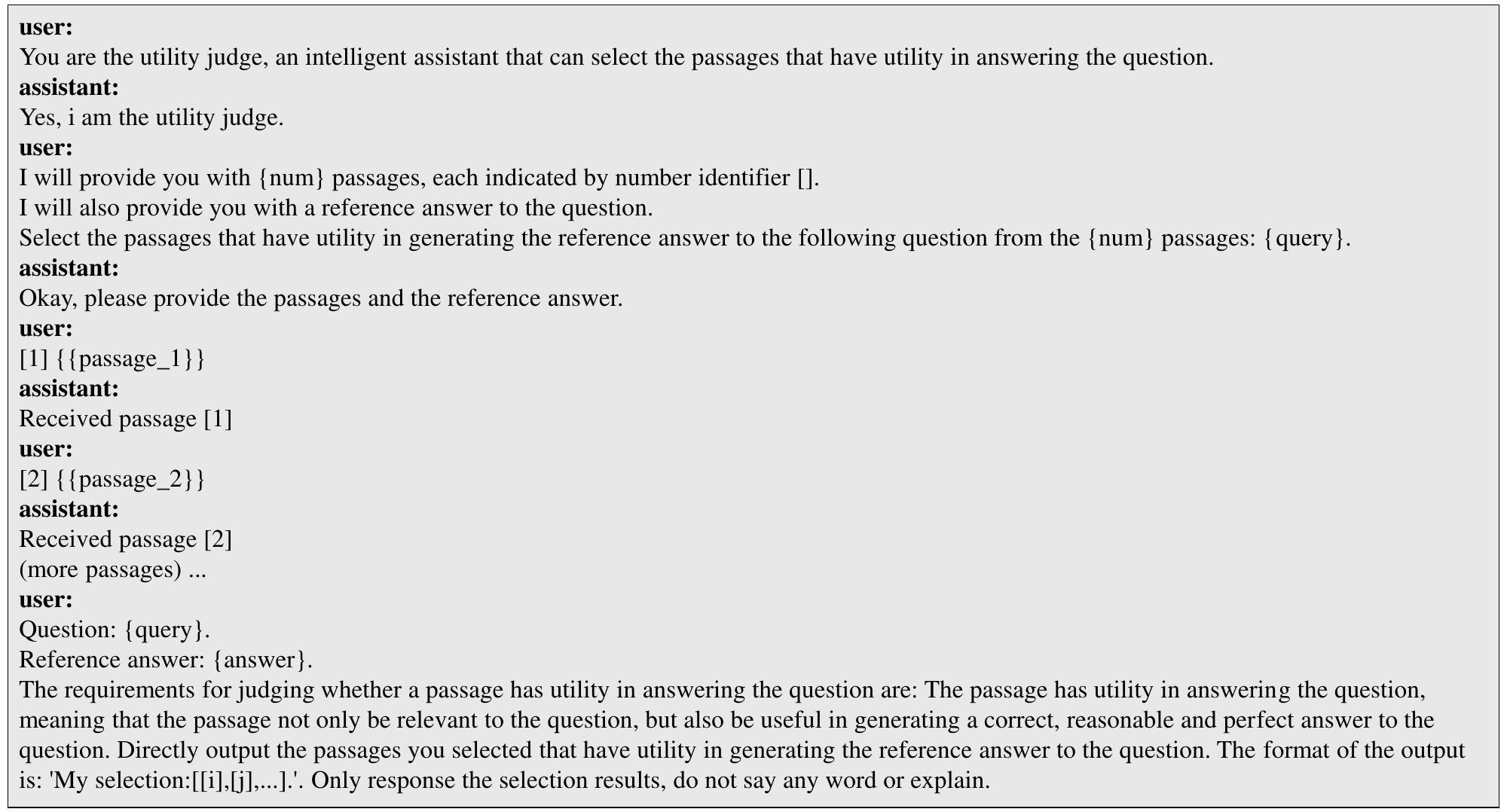}
    \caption{Instruction in the listwise approach.}
    \label{fig:listwise-set-prompt}
\end{figure*}

\begin{figure*}[t]
    \centering
    \includegraphics[width=\linewidth]{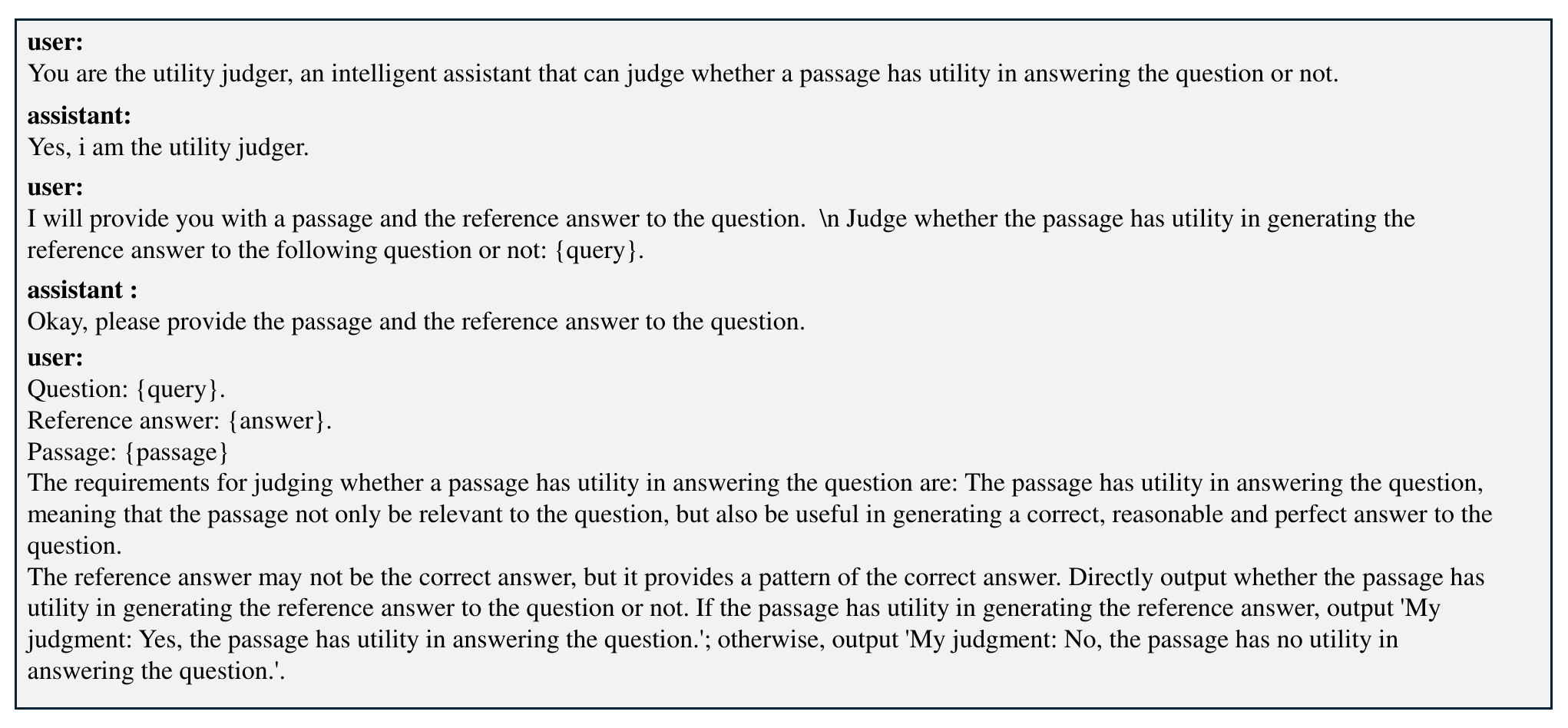}
    \caption{Instruction in the pointwise approach.}
    \label{fig:pointwise-prompt}
\end{figure*}

For the prompts of the NQ dataset using ChatGPT, we follow the setting of \citet{Zhang2024AreLL}, otherwise, we use the following prompts. 
Following \citet{sun2023chatgpt}, we input N passages using the form of multiple rounds of dialogue in the listwise approach. 
The prompts we used in our experiments are shown 
in Figure \ref{fig:listwise-set-prompt} and Figure \ref{fig:pointwise-prompt}.
\label{sec:app:instruction}

\subsection{Instruction of the Ranking Approach}
\label{app:sec:ranking}
\begin{figure*}[t]
    \centering
    \includegraphics[width=\linewidth]{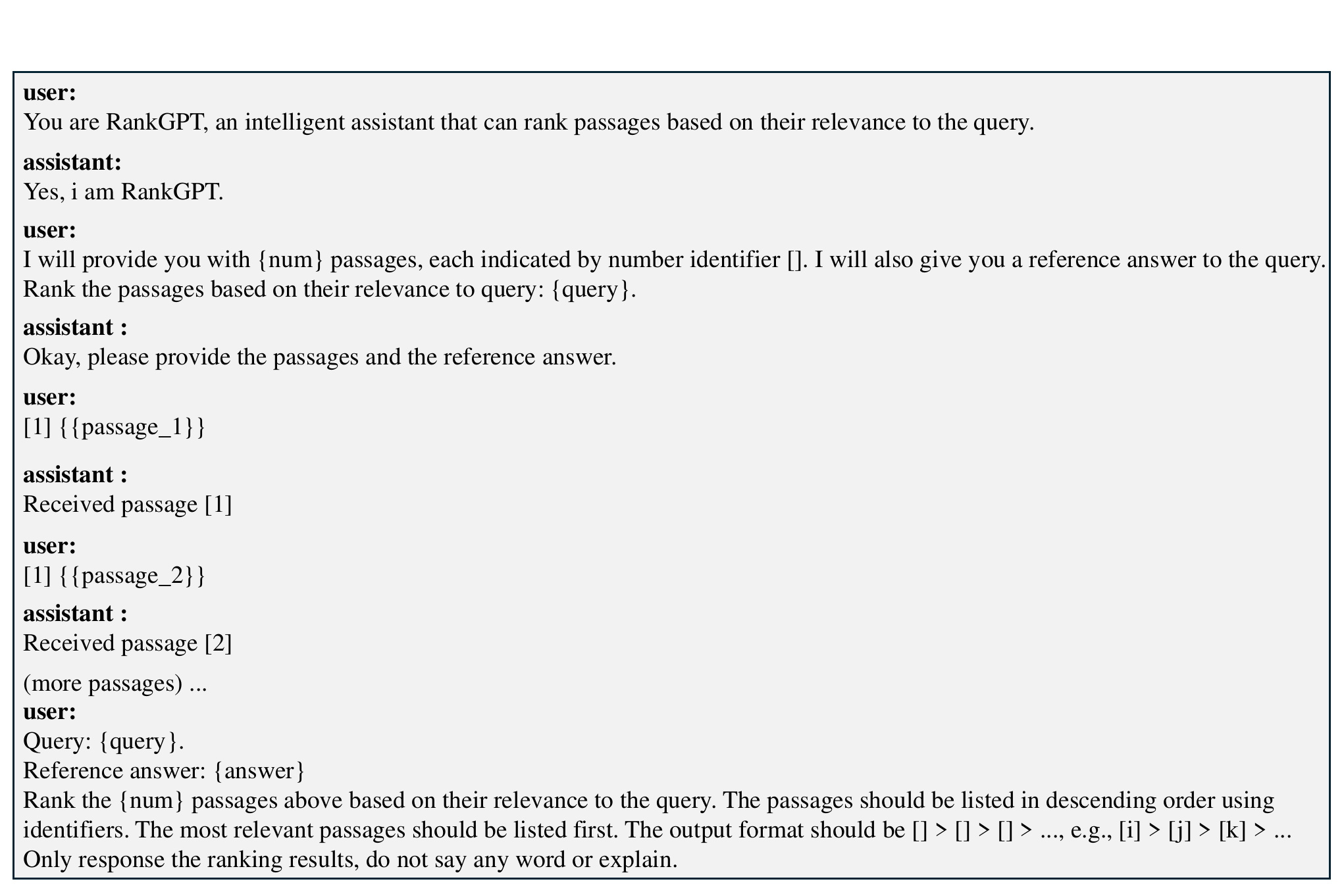}
    \caption{Instruction of the relevance ranking approach in our \modelname.}
    \label{fig:relevance-ranking}
\end{figure*}

\begin{figure*}[t]
    \centering
    \includegraphics[width=\linewidth]{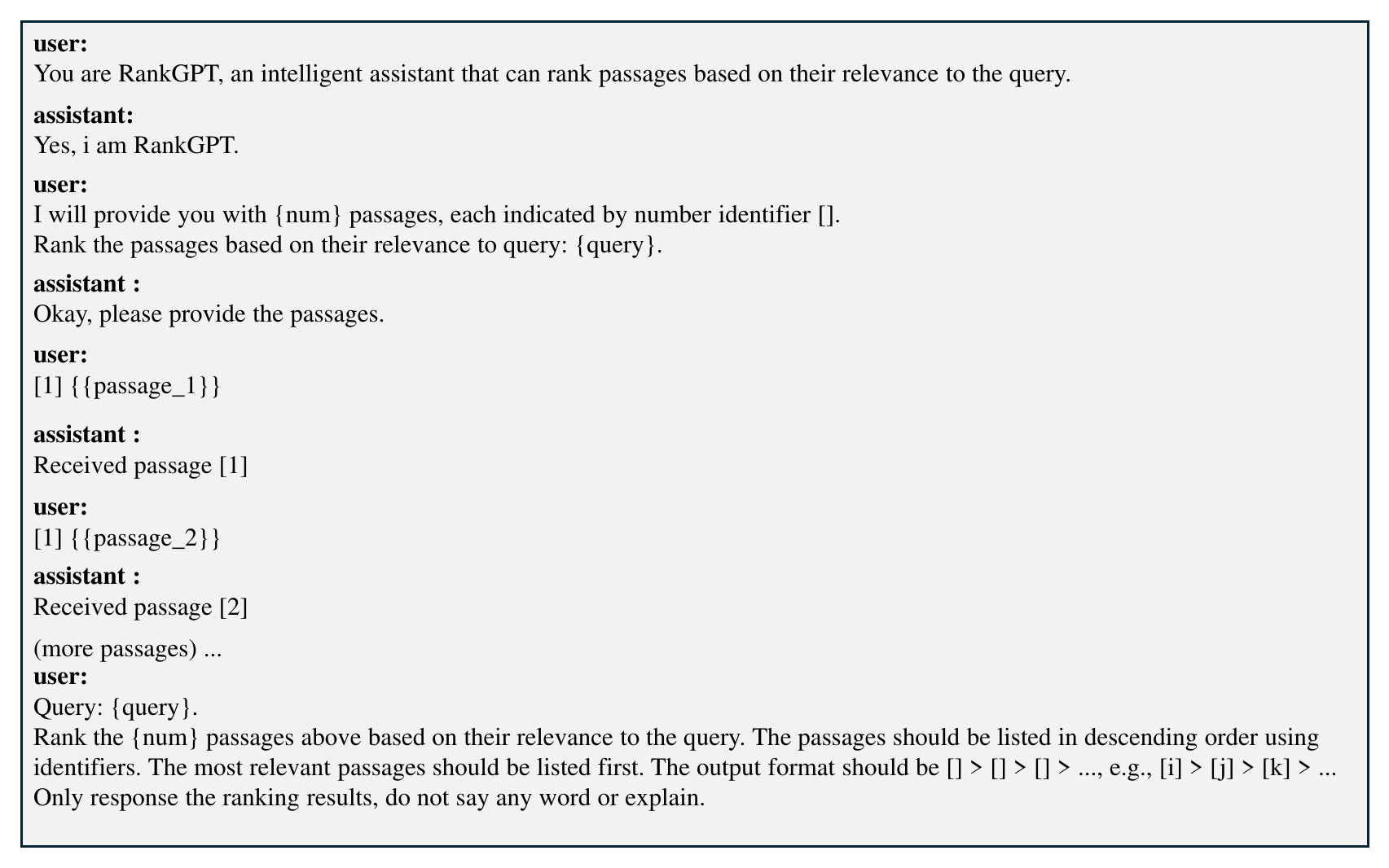}
    \caption{Instruction of the ranking approach in \citet{sun2023chatgpt}.}
    \label{fig:sun-rank-prompt}
\end{figure*}

For RankGPT, we directly use the instruction of \citet{sun2023chatgpt} for relevance ranking, as shown in Figure \ref{fig:sun-rank-prompt}. 
For the relevance ranking in our \modelname, the instructions are shown in Figure \ref{fig:relevance-ranking}.


\subsection{Instruction of Answer Generation}
\begin{figure*}[t]
    \centering
    \includegraphics[width=\linewidth]{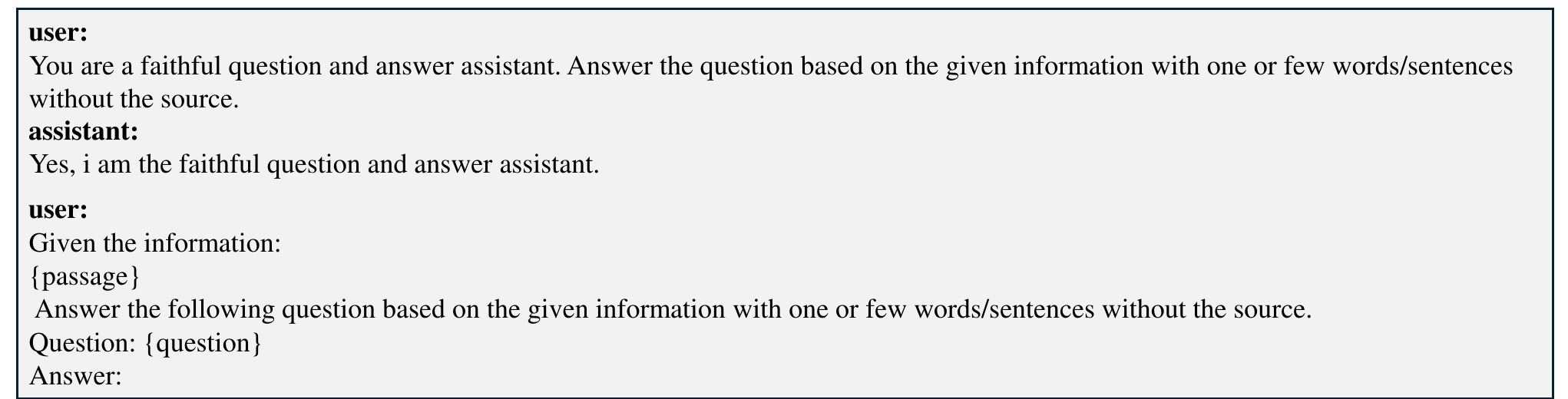}
    \caption{Instruction of the explicit answer generation.}
\vspace{-1mm}
    \label{fig:exp-answer-prompt}
\end{figure*}

\begin{figure*}[t]
    \centering
    \includegraphics[width=\linewidth]{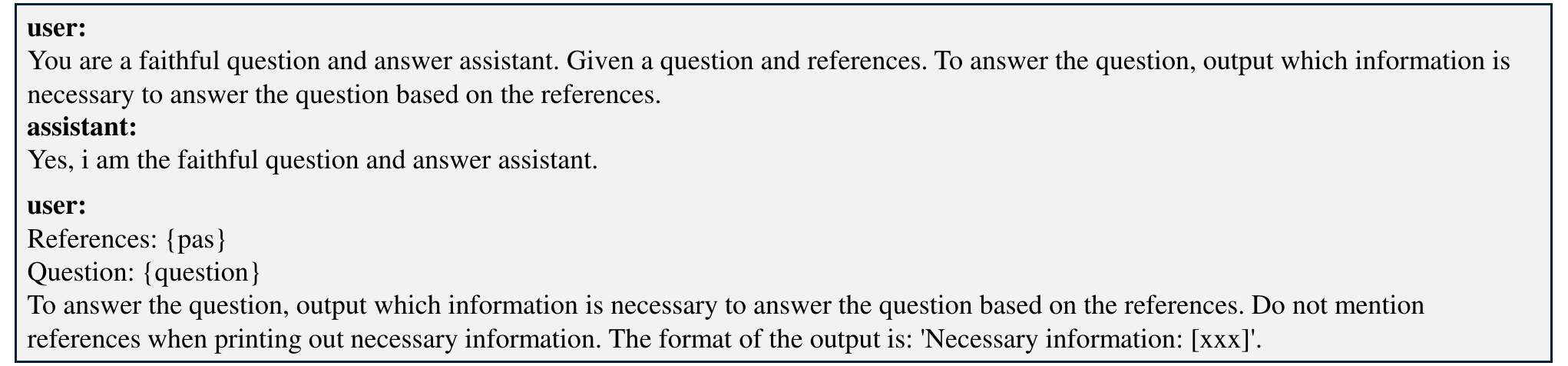}
    \caption{Instruction of the implicit answer generation.}
    \label{fig:imp-answer-prompt}
\end{figure*}

\citet{li2023llatrieval} utilize LLM to generate the missing information in the provided documents for the current question and then re-retrieve it as relevant feedback. 
Therefore, we have also designed two kinds of pseudo answers for utility judgments, i.e., 
\begin{enumerate*}[label=(\roman*)]
\item the explicit answer, which produces an answer based on the given information, and 
\item the implicit answer, which does not answer the question directly but gives the information necessary to answer the question.
\end{enumerate*}
The two instructions are shown in Figure \ref{fig:exp-answer-prompt} and Figure \ref{fig:imp-answer-prompt}. 

\end{document}